\documentclass[12pt]{article}
\usepackage{graphicx}
\usepackage{color}
\usepackage{cite}
\def\LLf{\hbox{${L}_f$}}

\def \beq{\begin{equation}}
\def \eeq{\end{equation}}
\def\eqref#1{(\ref{#1})}
\def\bea{\begin{eqnarray}}
\def\eea{\end{eqnarray}}
\def\jpsi{J\kern-0.15em/\kern-0.15em\psi\kern0.15em}

\def\URLtilde{\lower0.2em\hbox{$\tilde{\phantom{a}}$}}
\def\mycomm#1{\hfill\break\strut\kern-3em{\color{red}\tt ====> #1
\color{black}}\hfill\break}

\newcommand\tev{\mathrm{TeV}}
\newcommand\gev{\mathrm{GeV}}

\newcommand\bri{\mathcal{B}_i}
\newcommand\brf{\mathcal{B}_f}
\usepackage[colorlinks=true,linkcolor=darkblue,citecolor=darkblue]{hyperref}
\definecolor{darkblue}{RGB}{10,10,100}
\newcount\timecount
\newcount\hours \newcount\minutes  \newcount\temp \newcount\pmhours
\hours = \time
\divide\hours by 60
\temp = \hours
\multiply\temp by 60
\minutes = \time
\advance\minutes by -\temp
\def\hour{\the\hours}
\def\minute{\ifnum\minutes<10 0\the\minutes
\else\the\minutes\fi}
\def\clock{
\ifnum\hours=0 12:\minute\ AM
\else\ifnum\hours<12 \hour:\minute\ AM
\else\ifnum\hours=12 12:\minute\ PM
\else\ifnum\hours>12
\pmhours=\hours
\advance\pmhours by -12
\the\pmhours:\minute\ PM
\fi
\fi
\fi
\fi
}

\def\monthname{\relax\ifcase\month 0/\or January\or February\or
March\or April\or May\or June\or July\or August\or September\or
October\or November\or December\else\number\month/\fi}

\def\bold#1{\setbox0=\hbox{$#1$}     \kern-.025em\copy0\kern-\wd0
\kern.05em\copy0\kern-\wd0
\kern-.025em\raise.0433em\box0 }

\begin{document}
\setcounter{footnote}{1}
\rightline{EFI 15-1}
\rightline{TAUP 2990/15}
\vskip1.5cm

\centerline{\large \bf RADIATIVE RETURN CAPABILITIES OF A}
\centerline{\large \bf HIGH-ENERGY, HIGH-LUMINOSITY $e^+e^-$ COLLIDER}
\bigskip

\centerline{Marek Karliner$^a$\footnote{{\tt marek@proton.tau.ac.il}}.
    Matthew Low$^{b,c}$\footnote{{\tt mattlow@uchicago.edu}},
    Jonathan L. Rosner$^b$\footnote{{\tt rosner@hep.uchicago.edu}},
     and Lian-Tao Wang$^{b,c}$\footnote{{\tt liantaow@uchicago.edu}}}
\medskip

\centerline{$^a$ {\it School of Physics and Astronomy}}
\centerline{\it Raymond and Beverly Sackler Faculty of Exact Sciences}
\centerline{\it Tel Aviv University, Tel Aviv 69978, Israel}
\medskip

\centerline{$^b$ {\it Enrico Fermi Institute and Department of Physics}}
\centerline{\it University of Chicago, 5620 S. Ellis Avenue, Chicago, IL
60637, USA}
\medskip

\centerline{$^c$ {\it Kavli Institute for Cosmological Physics}}
\centerline{\it University of Chicago, 933 E. 56th Street, Chicago, IL
60637, USA}
\medskip

\begin{center}
ABSTRACT
\end{center}
\begin{quote}
An electron-positron collider operating at a center-of-mass energy $E_{CM}$ 
can collect events at all lower energies through initial-state radiation (ISR
or {\it radiative return}).  We explore the capabilities for radiative return
studies by a proposed high-luminosity collider at $E_{CM} = 250$ or 90 GeV, to
fill in gaps left by lower-energy colliders such as PEP, PETRA, TRISTAN, and
LEP.  These capabilities are compared with those of the lower-energy $e^+ e^-$
colliders as well as hadron colliders such as the Tevatron and the CERN
Large Hadron Collider (LHC).  Some examples of accessible questions in
dark photon searches and heavy flavor spectroscopy are given.
\end{quote}

\leftline{PACS codes: 13.66.Bc, 13.66.De, 13.66.Hk}

\section{Introduction \label{sec:intro}}

An electron-positron collider operating at a center-of-mass energy $E_{CM}$
can collect events at all lower energies through initial-state radiation (ISR).
This {\it radiative return} process has been used to good advantage in $e^+
e^-$ colliders such as DA$\Phi$NE, PEP-II, KEK-B, and LEP \cite{Denig:2006kj,%
Kluge:2008fb,Czyz:2010hj,Druzhinin:2011qd}. In the present paper we explore the
capabilities of a higher-energy high-luminosity $e^+e^-$ collider such as that
envisioned by CERN (FCC-ee) \cite{FCC} or China (CEPC) \cite{CEPC}, operating
at $E_{CM} \simeq 250$ or 90 GeV (functioning as a Giga- or Tera-$Z$ factory at
the latter energy) \cite{Ruan:2014xxa}, to perform radiative return studies of
physics at lower center-of-mass energies.

In order to fairly assess the capabilities of future colliders with past and
present colliders it is necessary to specify the total integrated luminosity
expected to be collected by future colliders.  Based on current design reports,
over 2 interaction points the CEPC is expected to collect 500 fb$^{-1}$ on the
$Z$ pole, which corresponds to approximately $1 \times 10^{10}$ $Z$'s, and 5
ab$^{-1}$ at $E_{CM} \simeq$ 250 GeV~\cite{CEPC:preCDR}.  The FCC-ee, over 4
interaction points, is expected to collect 50 ab$^{-1}$, which is roughly $1
\times 10^{12}$ $Z$'s, at $E_{CM} \simeq$ 90 GeV and 10 ab$^{-1}$ at $E_{CM}
\simeq$ 250 GeV~\cite{Gomez-Ceballos:2013zzn}.  Table~\ref{tab:futurelumis}
summarizes these numbers.  For convenience where the exact number of events is
not imporant, we shall occasionally quote results for a nominal integrated
luminosity of 1 ab$^{-1}$.  These values may be rescaled appropriately.

\begin{table}
\caption{Projected luminosities for the CEPC~\cite{CEPC:preCDR} and
FCC-ee~\cite{Gomez-Ceballos:2013zzn}.  These values are used throughout the
text.
\label{tab:futurelumis}}
\begin{center}
\begin{tabular}{c c c} \hline \hline
          & $\sqrt{s} =$ 90 GeV   & $\sqrt{s} =$ 250 GeV \\ \hline
 CEPC     & 0.5 ab$^{-1}$         &  5 ab$^{-1}$ \\
 FCC-ee   & 50 ab$^{-1}$         & 10 ab$^{-1}$ \\ \hline \hline
\end{tabular}
\end{center}
\end{table}

We briefly review some previous uses of radiative return in
Section~\ref{sec:previous}.  In Section~\ref{sec:resonance} we study narrow 
resonance production, while the discussion is extended to continuum production
in Section~\ref{sec:continuum}. 
Section~\ref{sec:benchmark} compares the reach of $e^+ e^-$ and hadron
colliders for two benchmark processes:  ``dark photon'' and $b \bar b$
production.  Some processes of interest in heavy flavor spectroscopy are noted
in Section~\ref{sec:flavor}.  We conclude in Section~\ref{sec:conclusions}.
Some calculational checks are contained in two Appendices:~\ref{sec:plumi}
and~\ref{sec:flumi}.

\section{Some previous uses of radiative return ~\label{sec:previous}}

\begin{figure}
\begin{center}
\includegraphics[width=0.6\textwidth]{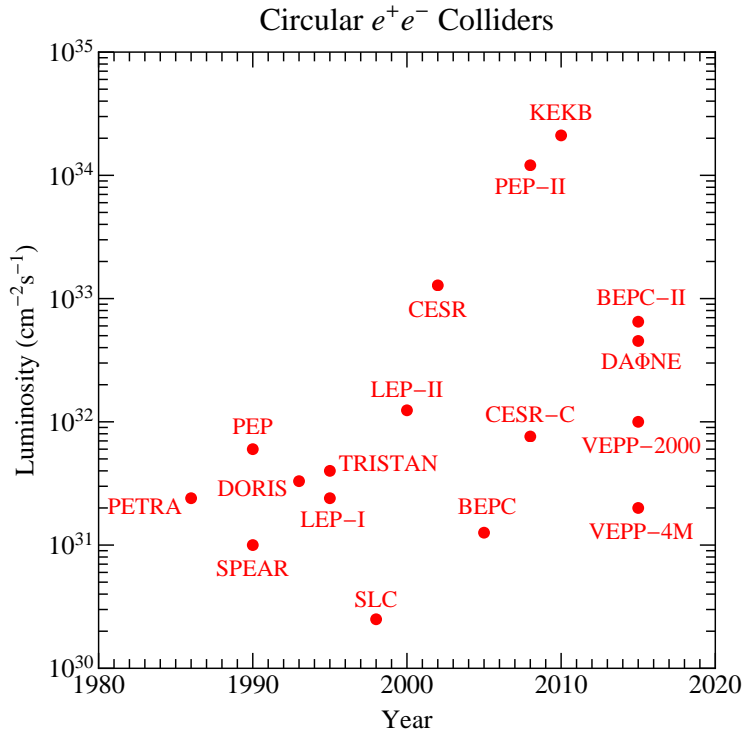}
\end{center}
\caption{Maximum instantaneous luminosities of circular $e^+ e^-$ colliders
versus time.  Adapted from Fig.~1 of Ref.~\cite{Biscari:2004fs}.
\label{fig:colls}}
\end{figure}

Considerable use has been made of radiative return in previous experiments
using electron-positron colliders.  In Table~\ref{tab:colls} we summarize
some parameters of experiments at these colliders \cite{PDG2014,Barnett:1996hr}.
Maximum instantaneous luminosities of circular $e^+e^-$ colliders are plotted
versus year in Fig.~\ref{fig:colls}.

\subsection{KLOE at DA$\Phi$NE}

The DA$\Phi$NE accelerator at Frascati operates near or at the CM energy
(1020 MeV) of the $\phi$ resonance.  It has studied the cross section $e^+ e^-
\to \pi^+ \pi^-$ at lower CM energies via the process $e^+ e^- \to \gamma
\pi^+ \pi^-$, where the photon is emitted in initial-state radiation, with the
main purpose of reducing the error in the hadronic vacuum polarization
contribution to the anomalous magnetic moment $a_\mu$ of the muon.  Three
sets of data are reported:  141.4 pb$^{-1}$ studying the interval $0.35
< M^2_{\pi\pi} < 0.95$ GeV$^2$ \cite{Aloisio:2004bu}, 240 pb$^{-1}$ studying
the same interval \cite{Ambrosino:2008aa,Babusci:2012rp}, and 230 pb$^{-1}$
studying $0.1 < M^2_{\pi\pi} < 0.85$ GeV$^2$ \cite{Ambrosino:2010bv}.  KLOE
also has searched for ``dark photons'' below 1 GeV decaying to $e^+e^-$ and
$\mu^+ \mu^-$, as noted in more detail later
\cite{Curciarello:2015xja,Palladino:2015jya}.

\subsection{CLEO at CESR}

The CLEO Collaboration used initial-state radiation to search for the
state $X(3872)$ \cite{Dobbs:2004di} in $e^+ e^-$ collisions at the
Cornell Electron Storage Ring (CESR).  The absence of a signal served as
partial evidence that the state did not have spin 1 and negative parity
and charge-conjugation eigenvalue.

\begin{table}
\caption{Instantaneous and/or integrated luminosities achieved at some $e^+e^-$
colliders.  Based in part on Section 30 of Ref.~\cite{PDG2014}, with values
from Ref.~\cite{Barnett:1996hr} for PETRA, PEP, and TRISTAN.  We thank G.
Alexander and S. L. Wu for help with some of these estimates.
\label{tab:colls}}
\begin{center}
\begin{tabular}{c c c c c} \hline \hline
Collider & Detector & CM energy & Max.\ ${\cal L}$ & $\int {\cal L} dt$ \\
         &          & (GeV)   & $(10^{30}$ cm$^{-2}$s$^{-1})$ & (fb$^{-1}$)
 \\ \hline
DA$\Phi$NE &  KLOE  & 1.02  & 453 & 2.5  \\
         &          & 1.00  & 453 & 0.23    \\
CESR     &  CLEO    & 9.46--11.30 & 1280 at 10.6 GeV & 15.1 \\
PEP-II   &  BaBar   & 10.58 & 12069 & 424.7   \\
         &          & 10.18 & $\ldots$ & 43.9    \\
KEK-B    &  Belle   & 9.46--10.89 & 21083 & 980 \\
PEP      &        &  29   &        60        & $1.167^a$ \\
PETRA    &        & 46.8$^b$  &  24 at 35 GeV  & $0.817^c$ \\
TRISTAN  &        & 64$^b$ &       40        & $0.942^d$ \\
LEP      &          & $M_Z$ & 24 & $0.808^e$ \\
         &          & $>130$ & 34--90  & $2.980^e$ \\ \hline \hline
\end{tabular}
\end{center}
\leftline{$^a$ Summed over detectors DELCO, HRS, MAC, Mark II, TPC/2$\gamma$}
\leftline{$^b$ Maximum value}
\leftline{$^c$ Summed over detectors CELLO, JADE, Mark J, PLUTO, TASSO}
\leftline{$^d$ Summed over detectors AMY, TOPAZ, VENUS}
\leftline{$^e$ Summed over detectors ALEPH, DELPHI, L3, OPAL}
\end{table}

\subsection{BaBar at PEP-II}

The initial-state radiation process has been used to great advantage by the
BaBar Collaboration at PEP-II.  Just in the past three years, papers have
appeared on the production of $\pi^+ \pi^- \pi^+ \pi^-$ \cite{Lees:2012cr};
$K^+K^-\pi^+\pi^-,~K^+K^-\pi^0\pi^0,~K^+K^-\pi^+\pi^-$ \cite{Lees:2011zi};
$\pi^+\pi^-$ \cite{Lees:2012cj}; $J/\psi \pi^+ \pi^-$ \cite{Lees:2012cn};
$p \bar p$ \cite{Lees:2013ebn}; $K^+ K^-$ \cite{Lees:2013gzt}; $\psi(2S)
\pi^+ \pi^-$ \cite{Lees:2012pv}; and a variety of final states with two
neutral kaons \cite{Lees:2014xsh}.  The CM energies and integrated luminosities
in Table~\ref{tab:colls} are those quoted for BaBar in the last paper.  The
final states involving light hadrons contribute to reducing the uncertainty on
the hadronic vacuum polarization contribution to $a_\mu$ and to the running
of the fine structure constant in precision electroweak studies, while those
involving $J/\psi$ and $\psi(2S)$ are of interest for resonant structures.

\subsection{Belle at KEK-B}

Since 2011 the Belle Collaboration has produced a couple of initial-state
radiation studies, involving production of $J/\psi K^+ K^-$ and $J/\psi K_S
K_S$ \cite{Shen:2014gdm}; and $\psi(2S) \pi^+ \pi^-$ \cite{Wang:2014hta}.
The focus of this work has been the search for resonant substructures in
the final states.

\subsection{LEP Collaborations}

The LEP entries in Table~\ref{tab:colls} refer to total integrated luminosities
at various energies.  Results from specific detectors are not always based on
these totals, as many of them were obtained before the full data sample was
available.

\subsubsection{ALEPH}

The reaction $e^+ e^- \to \gamma_{\rm ISR}\, \mu^+ \mu^-$ has been used by
the ALEPH Collaboration \cite{Barate:1997sp} to study the cross section and
forward-backward asymmetry for $e^+ e^- \to \mu^+ \mu^-$ in the CM
energy range 20--136 GeV.
 
\subsubsection{DELPHI}

The helicity structure in $e^+ e^- \to \mu^+ \mu^-$ is particularly sensitive
to new physics in the CM energy range around 80 GeV, which is accessible
through the process $e^+ e^- \to \gamma_{\rm ISR} \mu^+ \mu^-$  The DELPHI
Collaboration \cite{Abreu:1997uq} has studied this process, finding no
evidence for new physics.

\subsubsection{L3}

The L3 Collaboration has used ISR to measure muon pair production in $e^+ e^-$
collisions between 50 and 86 GeV \cite{Acciarri:1996wg}.  This study was
motivated in part by the need to fill a gap between the maximum energy of
the TRISTAN accelerator at KEK (about 62 GeV) and the mass of the $Z$.

\subsubsection{OPAL}

The OPAL Collaboration pioneered the use of ISR at LEP to fill the
aforementioned gap between 62 GeV and $M_Z$ \cite{Acton:1991dq}.  No deviations
from the standard model were found, albeit with a very early data sample.
The collaboration used radiative fermion pair events to perform a LEP
beam energy measurement \cite{Abbiendi:2004zw}.  Fig. 1 of this reference
gives an idea of the yield that may be expected from radiative return studies
with beam energies approaching 200 GeV.  An extensive study was made,
accumulating a total of 1.132 fb$^{-1}$ at LEP CM energies between 183 and 207
GeV.  Aside from a copious $Z$ peak, the numbers of events per 2 GeV subenergy
at a subenergy of 125 GeV were about 200 for $q \bar q$, a dozen for $\mu^+
\mu^-$, and three dozen for $\tau^+ \tau^-$.  Supposing one had a sample of
1 ab$^{-1}$ at $E_{CM} = 250$ GeV, one might expect ${\cal O}(2 \times 10^5)$
hadronic events, ${\cal O}(10^4)~\mu^+\mu^-$ events, and a few tens of thousand
$\tau^+ \tau^-$ events, per 2 GeV bin in invariant mass around 125 GeV.

\section{Resonance production \label{sec:resonance}}

\begin{figure}
\begin{center}
\includegraphics[width=0.46\textwidth]{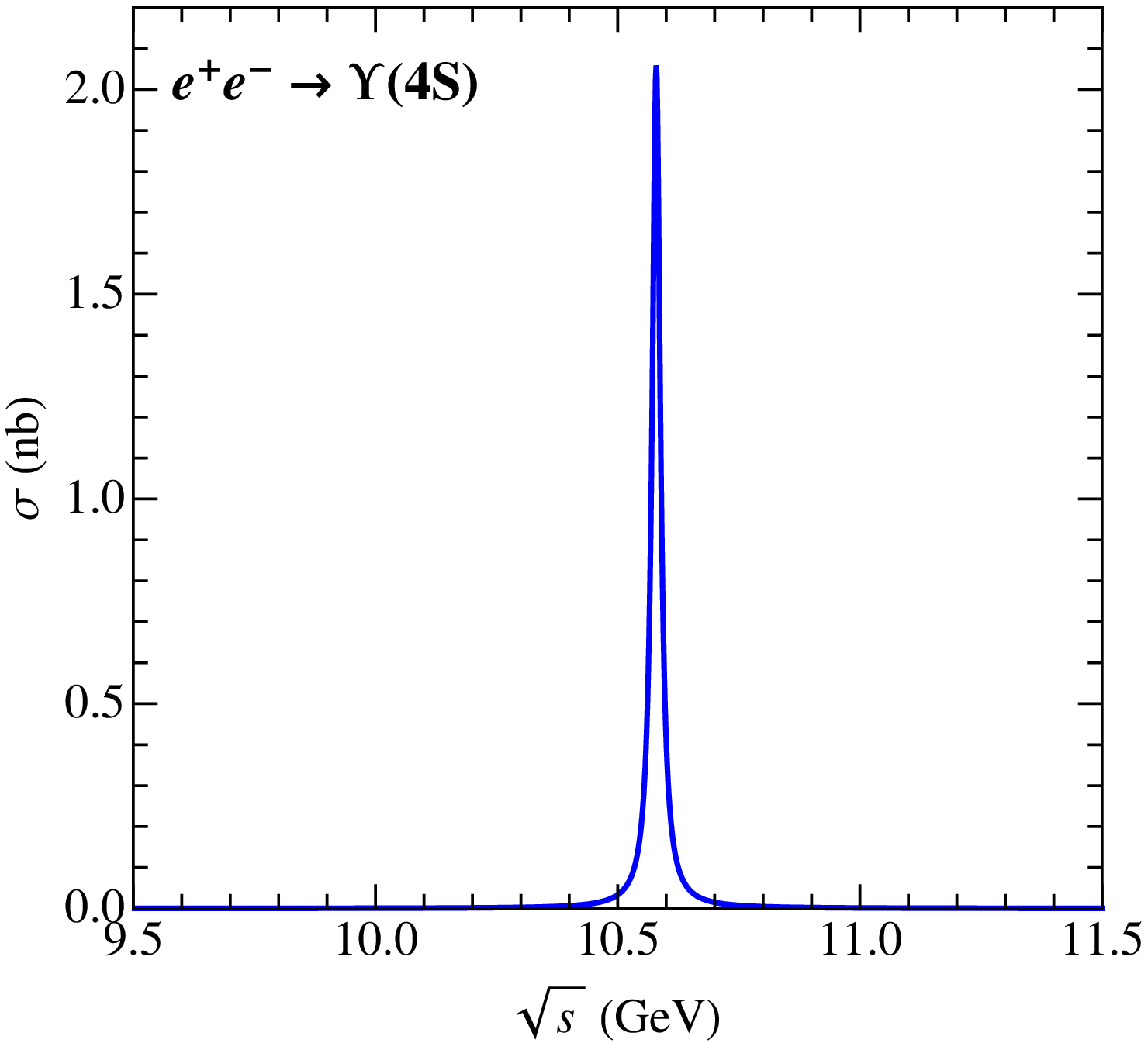}
\includegraphics[width=0.46\textwidth]{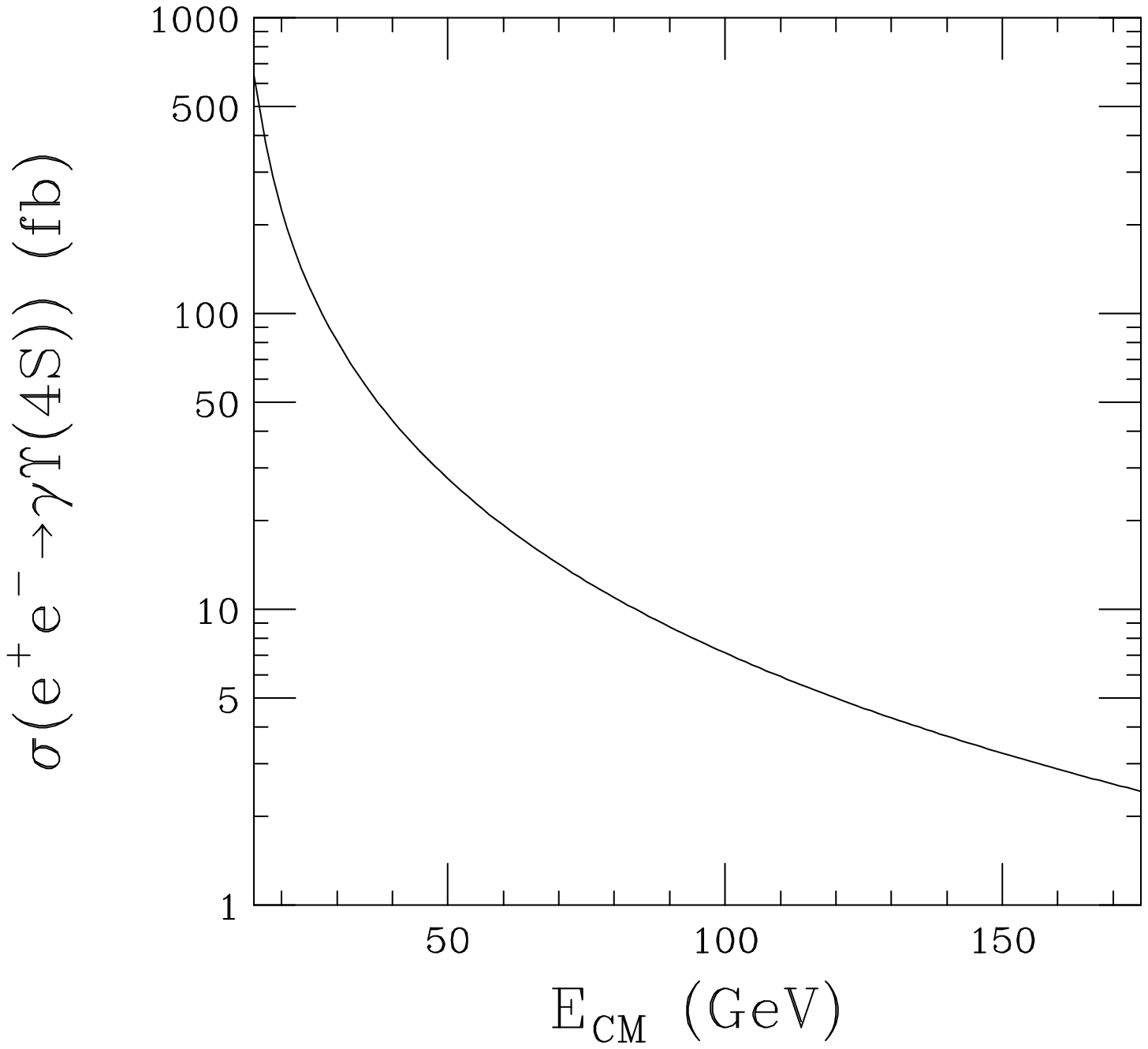}
\end{center}
\caption{Cross section for $e^+ e^- \to \Upsilon(4S)$ (left) and including
the emission of a photon at an $e^+e^-$ collider with CM energy $E_{CM}$
(right).
\label{fig:4s}}
\end{figure}

The cross section for electron-positron production of a vector meson resonance
$R$ with mass $m_R$ and $e^+ e^-$ partial width $\Gamma_{ee}$ decaying to a
final state $f$ with partial width $\Gamma_f$ may be written near resonance as
\beq \label{eqn:xsec}
\sigma(e^+ e^- \to R \to f;~s) = \frac{12 \pi \Gamma_{ee} \Gamma_f}
 {(s-m_R^2)^2+ (m_R \Gamma_R)^2}~~,
\eeq
where $s = E_{CM}^2$, and $m_R$ and $\Gamma_R$ are the resonance mass and total
width.\footnote{An extensive discussion of possible modifications of this
expression, including multiplication of $\Gamma$ by the factor $s/m^2_R$ to
ensure $1/s$ behavior of the cross section at high $s$, is given
in Ref.~\cite{modifs}.}
For the $\Upsilon(4S)$, whose decays are almost exclusively to $B \bar
B$ final states, the leptonic branching ratio is quoted by the Particle Data
Group \cite{PDG2014} as $1.57 \times 10^{-5}$ while the total width is 20.5 MeV,
leading to a leptonic partial width $\Gamma_{ee} = 0.322$ keV. We shall use this
value, noting that it is mildly inconsistent with the Particle Data Group's
average of 0.272 keV.  The mass is $10.5794 \pm 0.0012$ GeV; the cross section
at the resonance peak is about 2.06 nb.  The resonance shape is shown at the
left in Fig.~\ref{fig:4s}.

A resonance $R$ may be produced by the radiative return process $e^+ e^- \to
\gamma R$, where the electron or positron of beam energy $E = E_{CM}/2$
radiates a fraction $1-x$ of its energy and is left with energy $xE$.
Neglecting the small electron mass, the squared effective mass of the $e^+
e^-$ system is then $xs$.  An electron beam of energy $E$ radiates a photon
and ends up with an energy $xE$ with a probability
per unit $x$ \cite{Chen:1975sh} denoted by
\beq \label{eqn:split}
f_e(x,\sqrt{s},p_{T,{\rm cut}}) = \frac{\alpha}{\pi} \frac{1+x^2}{1-x}
 \ln \frac{E}{p_{T,{\rm cut}}}~,
\eeq
where the minimum photon transverse momentum $p_{T,{\rm cut}}$ provides a
collinear cutoff.\footnote{The numerator of the logarithm is sometimes taken to
be $2E = \sqrt{s}$.}
In the absence of an explicit choice of cutoff, it is provided by the electron
mass $m_e$, which we shall use in much of what follows.  The cross section for
production of the resonance $R$ by radiative return, where $R$ decays to the
final state $f$, is then
\beq
\sigma(e^+ e^- \to \gamma R \to \gamma f) = \frac{2 \alpha}{\pi} \ln
\frac{E}{m_e} \int_0^1 dx \frac{1+x^2}{1-x} \sigma(e^+ e^- \to R \to f;~xs)~~,
\eeq
where the factor of two comes from the fact that either lepton can radiate
the photon.  In the narrow-resonance approximation, the integral in this
expression can be done in closed form, with the result
\beq \label{eqn:sigrr}
\sigma(e^+ e^- \to \gamma R \to \gamma f) \simeq 24 \alpha\pi \ln \frac{E}{m_e}
~\frac{1+x_0^2}{1-x_0}~\frac{\Gamma_{ee}{\cal B}_f}{m_R~s}~~,
\eeq
where $x_0 = m_R^2/s$ and ${\cal B}_f=\Gamma_f/\Gamma_R$ denotes the branching
fraction into the final state $f$.  The cross section for $e^+e^- \to
\Upsilon(4S)$ including the emission of a photon is shown as a function of
$e^+e^-$ CM energy in Fig.~\ref{fig:4s} (right).

The proposed high-energy electron-positron colliders at CERN and in China
anticipate integrated luminosities of 50 ab$^{-1}$ and 0.5 ab$^{-1}$, respectively, at
CM energy 90 GeV, and 10 ab$^{-1}$ and 5 ab$^{-1}$, respectively, at 250 GeV~\cite{CEPC:preCDR,Gomez-Ceballos:2013zzn}.
The observation of a new resonance with at least 10 events would then require cross sections
of at least 0.2 and 20 ab at CERN or China, respectively, at 90 GeV, or
at least 1 and 2 ab, respectively, at 250 GeV.

\begin{figure}
\begin{center}
\includegraphics[width=0.65\textwidth]{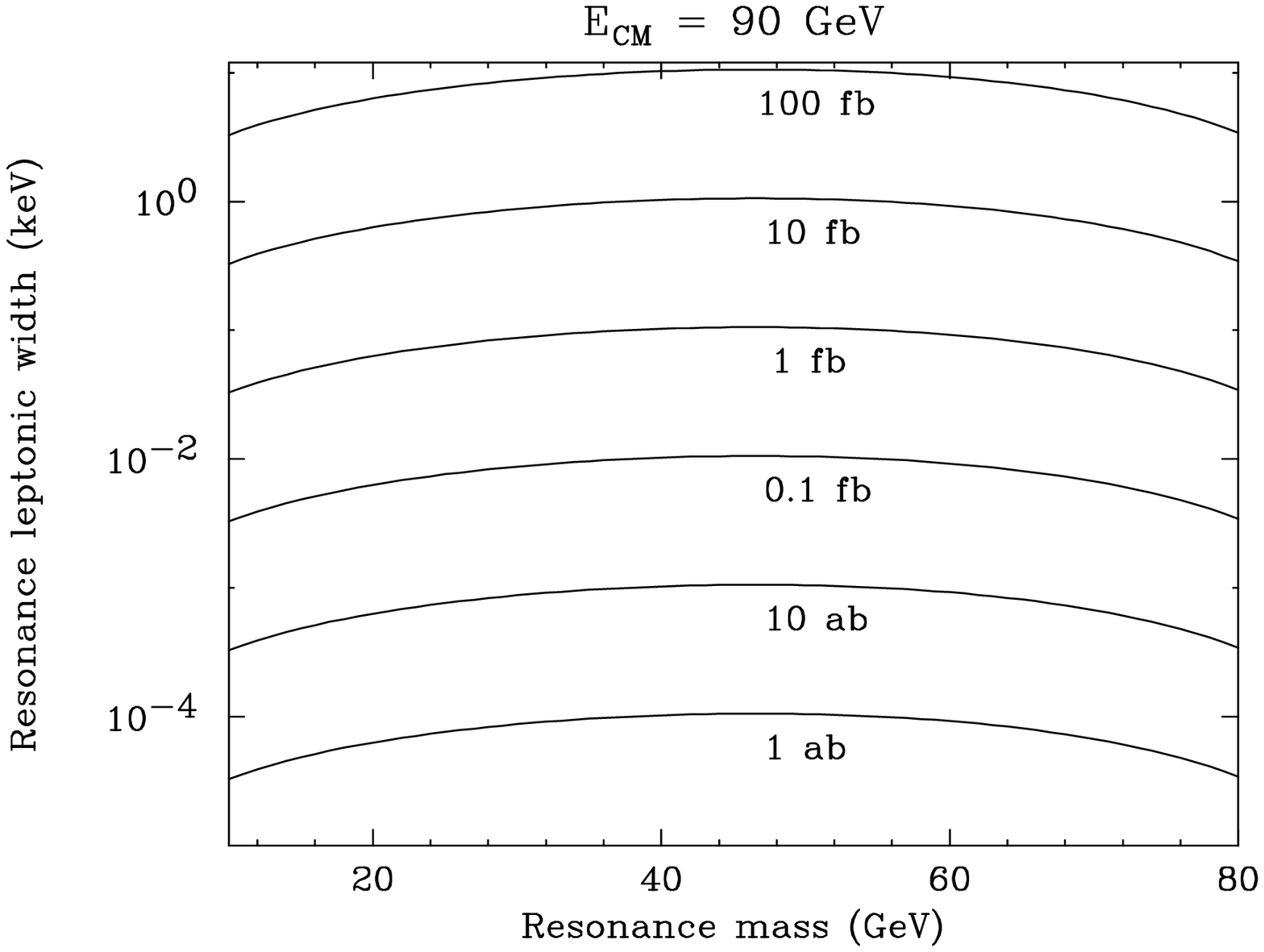}
\includegraphics[width=0.65\textwidth]{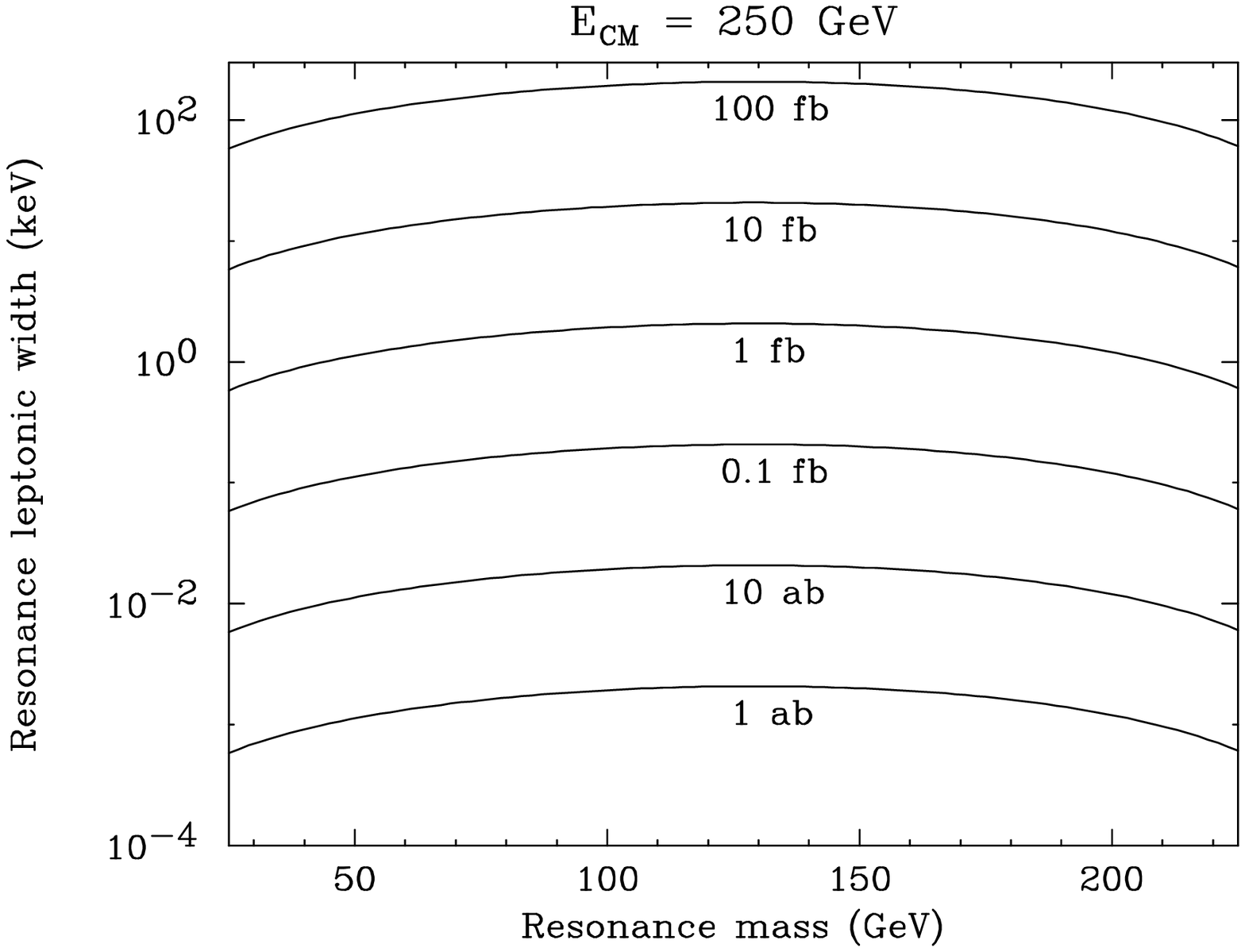}
\caption{Contours of equal cross section for radiative return production
of a resonance with leptonic width $\Gamma_{ee}$ (assuming 100\%
branching fraction to a final state $f$).  Top:  $E_{CM} = 90$ GeV; bottom:
$E_{CM} = 250$ GeV.
\label{fig:scan}}
\end{center}
\end{figure}

Fig.~\ref{fig:scan} illustrates contours of equal cross section for an
$e^+e^-$ collider with CM energy 90 (top) and 250 (bottom) GeV to produce a
resonance of mass $m_R$ via radiative return.  These results imply a cross
section of 9.17 fb for the $\Upsilon(4S)$ produced by radiative return at
$E_{CM} = 90$ GeV, given an assumed leptonic partial width of $\Gamma_{ee} =
0.322$ keV~\cite{PDG2014}.  For a given $E_{CM}$, the lowest sensitivity
appears to occur for a resonance mass roughly equal to $E_{CM}/2$, i.e., the
beam energy.

\begin{figure}
\begin{center}
\includegraphics[width=0.7\textwidth,{angle=90}]{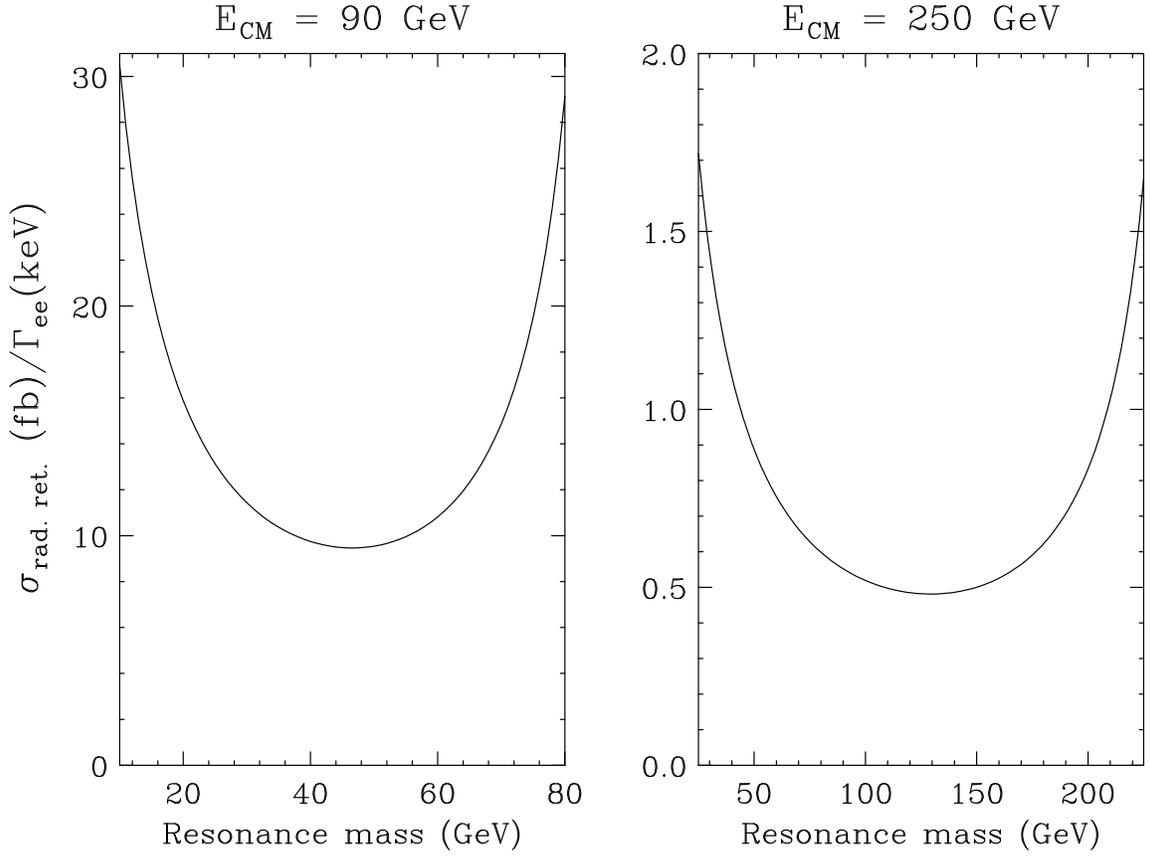}
\caption{$\sigma(e^+ e^- \to \gamma R \to \gamma f)/\Gamma_{ee}{\cal B}_f$ as
a function of resonance mass for $E_{CM} = 90$ (left) and 250 (right) GeV.
\label{fig:ratio}}
\end{center}
\end{figure}

The results of Fig.~\ref{fig:scan} can be expressed in more universal form.
In the narrow-resonance approximation the predicted radiative return cross
section, Eq.~\eqref{eqn:sigrr}, is directly proportional to $\Gamma_{ee} {\cal
B}_f$, so the ratio $\sigma(e^+ e^- \to \gamma R \to \gamma f)/\Gamma_{ee}
{\cal B}_f$ is a function only of $s$ and $m_R$.  In Fig.~\ref{fig:ratio} we
plot this ratio as a function of resonance mass for two values of $E_{CM}$.

The leptonic widths of known and fictitious quarkonium states can serve as
benchmarks for the interpretation of Figs.~\ref{fig:scan} and \ref{fig:ratio}.
They are summarized for 1S states in Table~\ref{tab:lep} \cite{PDG2014,%
Moxhay:1984rt}.  (The bound state of an actual top quark $t$ of mass $\sim 173$
GeV/$c^2$ and a $\bar t$ is highly unstable due to the weak decay of the $t$
or $\bar t$.)

\begin{table}
\caption{Leptonic widths of known and fictitious 1S quarkonium states.
\label{tab:lep}}
\begin{center}
\begin{tabular}{c c c c c} \hline \hline
   1S   &  Quark &    Quark   & $\Gamma_{ee}$ & Ref.\ \\
 state  & charge & mass (GeV) & (keV) & \\ \hline
$J/\psi$ & 2/3 & 1.4 & $5.55\pm0.14\pm0.02$ & \cite{PDG2014} \\
$\Upsilon$ & --1/3 & 4.8 & $1.340\pm0.018$ & \cite{PDG2014} \\
Toponium     & 2/3 & 40 & 6.5 & \cite{Moxhay:1984rt} \\
(fictitious) & 2/3 & 45 & 6.7 & \cite{Moxhay:1984rt} \\ \hline \hline
\end{tabular}
\end{center}
\end{table}

The leptonic width of an S-wave quarkonium $Q \bar Q$ bound state of mass $M$
is given by \cite{VanRoyen:1967nq}
\beq
\Gamma_{ee}(Q \bar Q) = \frac{16 \pi \alpha^2e_Q^2}{M^2} |\Psi(0)|^2~,
\eeq
where $e_Q$ is the quark charge, $|\Psi(0)|^2$ is the square of the wave
function at the origin, and we have neglected relativistic and QCD corrections.
One sees that the leptonic widths in Table~\ref{tab:lep} are governed primarily 
by the square of the corresponding quark charge.  Over a wide range of quark 
mass, the decrease of the $1/M^2$ factor is approximately compensated by a 
corresponding growth of $|\Psi(0)|^2$.

Note that for weakly coupled resonances typically the width is $\propto M$,
which needs to be accounted for when interpreting Figs.~\ref{fig:scan}
and~\ref{fig:ratio}.  In this case the factor of $\Gamma_{ee}/M$ in
Eq.~\eqref{eqn:sigrr} is mostly independent of mass and the cross
section is governed purely by the factor $(1+x_0^2)/(1-x_0)$, falling
monotonically with decreasing resonance mass.

This discussion is oversimplified because it neglects the off-shell
process $e^+ e^- \to \gamma Z^* \to \gamma R$, in the absence of assumptions
about how $R$ couples to $Z$.  However, it gives an idea of the orders of
magnitude necessary to find a previously missed resonance in the mass
range below that in which the $Z^*$ contributes appreciably (e.g., below
about 60 GeV, the CM energy accessible to TRISTAN).

\section{Continuum production \label{sec:continuum}}

\begin{figure}
\begin{center}
\includegraphics[width=0.48\textwidth]{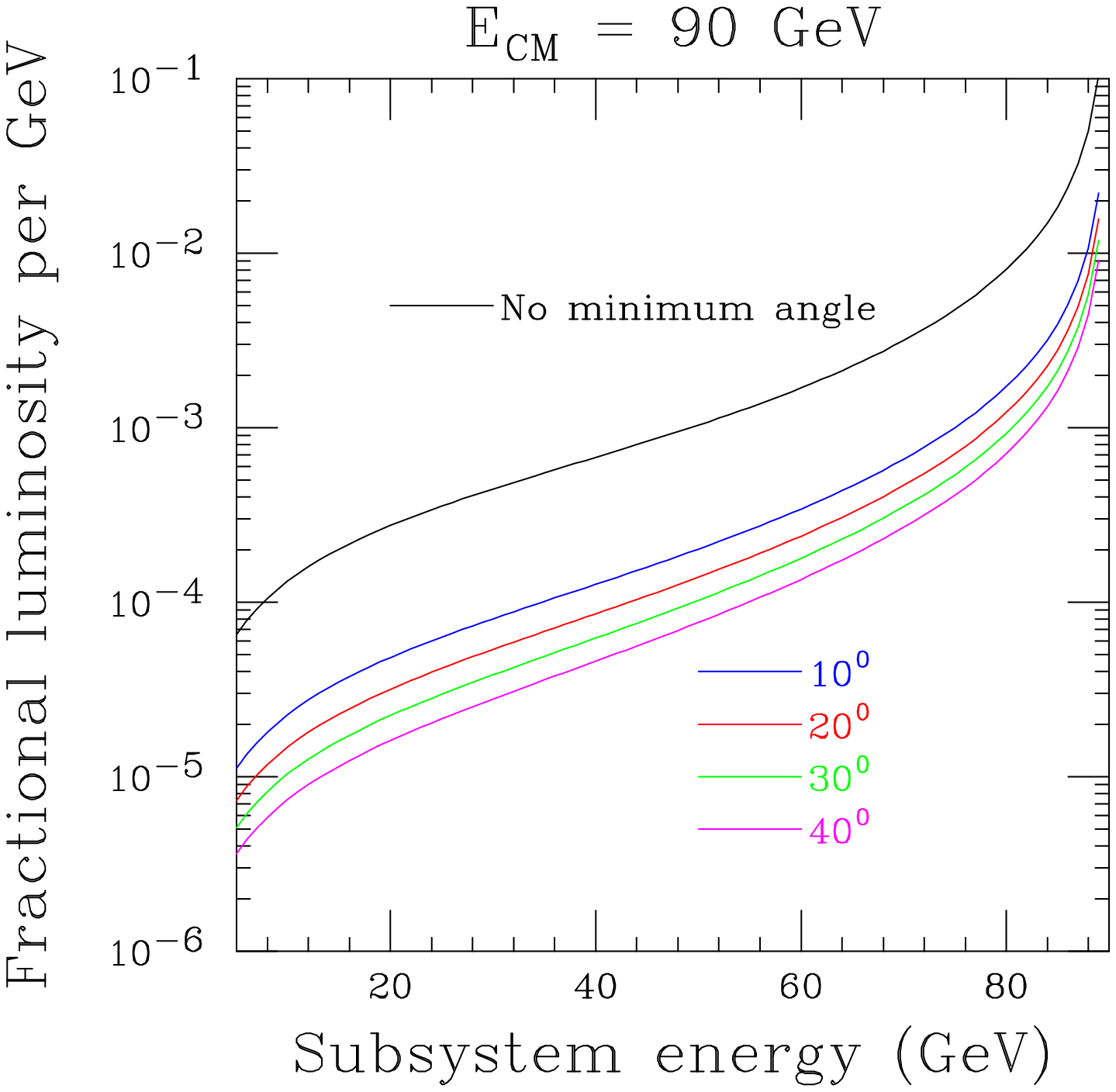}
\includegraphics[width=0.48\textwidth]{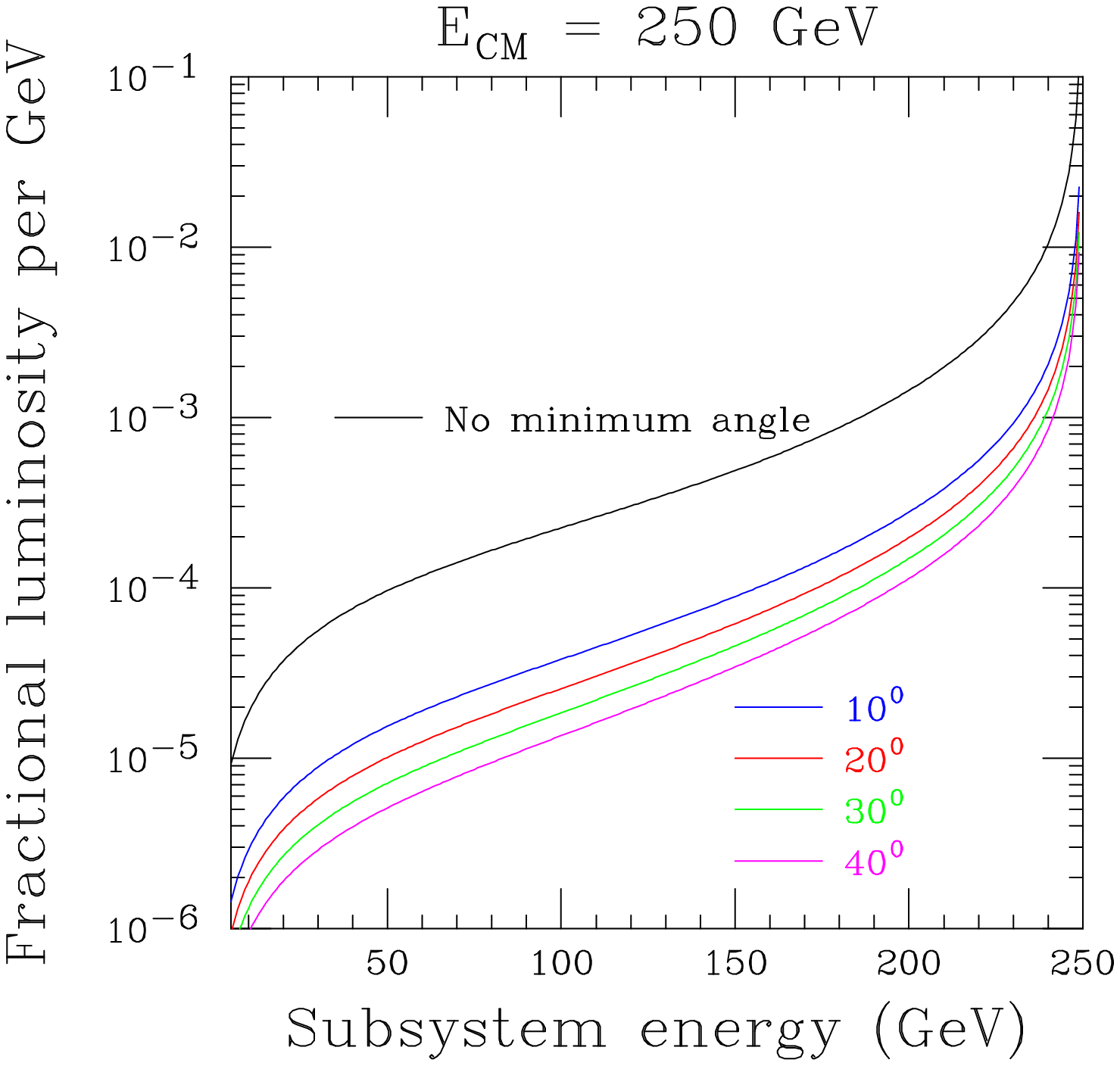}
\end{center}
\caption{Fractional luminosity \LLf\ as a function of subsystem energy
$\hat E_{CM}$ for $E_{CM} = 90$ (left) and 250 (right) GeV.  Top curves:  No
minimum angle; infrared cutoff provided by $\ln(E/m_e)$ [Eq.~\eqref{eqn:fl}].
Lower curves, top to bottom:  $\theta_0 = 10,~20,~30,~40^\circ$.
\label{fig:frac}}
\end{figure}

An important quantity is the effective luminosity of a high-energy collider
for studying any given process at lower center-of-mass energy.  Defining
$\sigma(s) \equiv \sigma(e^+ e^- \to \gamma f;s)$ and $\hat \sigma(\hat s)
\equiv \sigma(e^+e^- \to f;\hat s)$ for a given final state $f$, the relation
between the two is
\beq
\frac{d \sigma(s)}{dx} = \frac{2 \alpha}{\pi} \frac{1+x^2}{1-x}
 \ln \frac{E}{m_e} \hat\sigma(\hat s)~,
\eeq
where $x = \hat s/s$.  The subsystem CM energy may be denoted $\hat E_{CM}
= \sqrt{\hat s}$.  The cross section per unit $\hat E_{CM}$ times an
interval $\Delta$ of $\hat E_{CM}$ is then
\bea
\frac{d \sigma(s)}{d \hat E_{CM}} \Delta & = & \frac{4 \alpha \hat E_{CM}}
{\pi s}~\frac{1+x^2}{1-x}~\Delta~\ln \frac{E}{m_e}~ \hat \sigma(\hat s)
 \nonumber \\
 & \equiv & \LLf \hat \sigma(\hat s)~, \label{eqn:fl}
\eea
where \LLf\ is the {\em fractional luminosity} per $\hat E_{CM}$ bin 
of size $\Delta$.
Examples of this function for a bin width of $\Delta = 1$ GeV are shown in
the top curves of Fig.~\ref{fig:frac}.

For low $\hat E_{CM}$ one may take $(1+x^2)/(1-x) \simeq 1$ in Eq.~\eqref{eqn:fl}.
Integrating from $\hat E_{CM}^{\rm min}=10$ GeV to $\hat
E_{CM}^{\rm max} = 30$ GeV, one then finds
\beq
\LLf = \frac{2 \alpha}{\pi s}[(\hat E_{CM}^{\rm max})^2
 - (\hat E_{CM}^{\rm min})^2]\ln \frac{E}{m_e}~
\eeq
For $E_{CM} = (90,250)$ GeV we find $\LLf = (5.22,0.74) \times 10^{-3}$.
For a total of 1 ab$^{-1}$ at $E_{CM} = (90,250)$ GeV this then provides
a total integrated luminosity of (5220,740) pb$^{-1}$ in the range $10 \le
\hat E_{CM} \le 30$ GeV.  This exceeds integrated
luminosities at PEP or PETRA (see Table~\ref{tab:colls}).
 
\begin{figure}
\begin{center}
\includegraphics[width=0.90\textwidth]{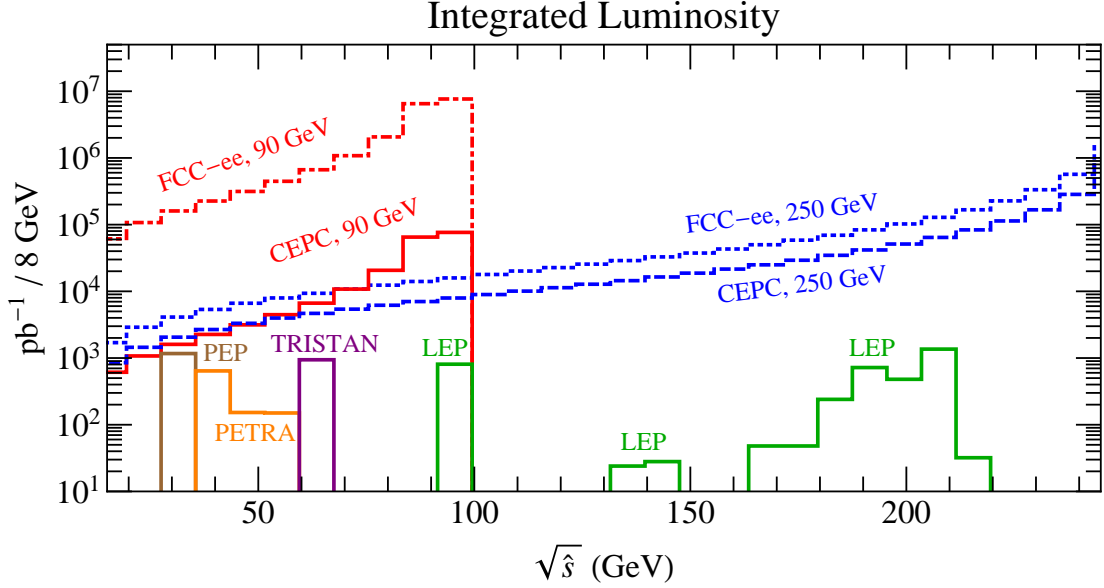}
\\\
\vskip-1.5cm
\strut
\end{center}
\caption{Integrated luminosity from past low energy $e^+ e^-$ colliders at their
nominal center-of-mass energies compared to the effective luminosity through
radiative return from future $e^+ e^-$ colliders at $\sqrt{s} =$ 90 or 250 GeV
(no minimum angle; see Fig.\ \ref{fig:frac} for effects of minimum angles).
The FCC-ee curves assume an integrated luminosity of 50 ab$^{-1}$ at 90 GeV and
10 ab$^{-1}$ at 250 GeV.  The CEPC curves assume an integrated luminosity of
0.5 ab$^{-1}$ at 90 GeV and 5 ab$^{-1}$ at 250 GeV.  Integrated luminosities of
PEP-II and Belle (Table \ref{tab:colls}) exceed those achievable
by radiative return at FCC-ee or CEPC running at 90 or 250 GeV.
\label{fig:summary}}
\end{figure}

Given the concept of fractional luminosity we can compute the effective luminosity
gathered at each center-of-mass energy via radiative return.  This is shown in
Fig.~\ref{fig:summary} for $\sqrt{s} = $ 90 and 250 GeV compared to the luminosity
collected directly at various other colliders.  In this figure it is illustrated 
clearly that a high-luminosity high-energy $e^+ e^-$ collider both competes with
and fills in gaps left by previous colliders.

Cleaner signals for radiative return may be obtained at the expense of
recorded events by demanding that the radiated photon make a minimum angle
$\theta_0$ with respect to the beam axis.  Let $\theta$ be the polar angle
of the radiated photon, and $z\equiv \cos \theta$, $z_0 \equiv \cos \theta_0$.
Using Eq.~(8) of \cite{Chen:1974wv}, we find the angular distribution
of the radiated photon for $m_e=0$ is given by
\beq
\frac{d^2 \sigma(s)}{d \hat E_{CM} \kern0.05em dz} = 
\frac{4 \alpha \hat E_{CM}}{\pi s}~\frac{\hat \sigma(\hat s)}{1-z^2} \left[ \frac{1-x}{4}(1+z^2)
+ \frac{x}{1-x} \right]~.
\eeq
This may be integrated between the desired limits of $\theta$, with the result
\beq \label{eqn:fla}
\int_{-z_0}^{z_0} dz \frac{d^2 \sigma(s)}{d \hat E_{CM} \kern0.05em dz} = \frac{4 \alpha
\hat E_{CM}}
{\pi s} \hat \sigma(\hat s) \left[ \frac{1-x}{2} \left(\ln \frac{1+z_0}{1-z_0}
 - z_0 \right) + \frac{x}{1-x}\ln \frac{1+z_0}{1-z_0} \right]~.
\eeq
The ratio between the left-hand side and $\hat \sigma(\hat s)$ is again a
fractional luminosity and is shown by the lower curves in Fig.~\ref{fig:frac},
again for a bin width of 1 GeV.  In the limit of small $\theta_0 \simeq
p_T/E_\gamma$, the leading-logarithmic term of Eq.~\eqref{eqn:fla} reduces to
the form in Eq.~\eqref{eqn:fl}.

Note that our computation of fractional luminosity utilizes factorization in the collinear limit.  In the Appendix, we perform the exact calculation for the
process $e^+ e^- \to \mu^+ \mu^-$ and find good agreement with our results in
Fig.~\ref{fig:frac}.

\section{Sensitivities of colliders for benchmark processes \label{sec:benchmark}}
\setcounter{footnote}{1}

We estimate the reach of radiative return studies using electron-positron
colliders for two benchmark processes: ``dark photon'' searches and $b \bar b$
production.  We compare these sensitivities with those of hadron colliders.
Although the latter have an advantage in total rate, it can only be realized
with considerable background suppression, such as provided for $b \bar b$
production by the VErtex LOcator (VELO) in the LHCb experiment.

\subsection{Dark photon search}

In this section we compute the reach for dark photons using radiative return as
a concrete example of a search for weakly-coupled resonances.  This has been
previously computed for GeV-scale dark photons using low energies colliders
like PEP-II in~\cite{Reece:2009un}.  In this work we focus on the 10's to 100's
of GeV scale, as discussed in~\cite{Curtin:2014cca}.  Other relevant work
includes~\cite{Pospelov:2007mp,ArkaniHamed:2008qp,Pospelov:2008jd,%
Pospelov:2008zw,Baumgart:2009tn,Cheung:2009qd,Batell:2009yf,Essig:2009nc,Hook:2010tw}.

For simplicity, we assume that a ``dark photon,'' denoted by $Z'$, is
kinetically mixed with a hypercharge gauge boson with amplitude $\epsilon$
\beq
  \mathcal{L} = -\frac{1}{4} \hat{B}_{\mu\nu }^2 -\frac{1}{4}\hat{Z}_{\mu\nu}'^2
  + \epsilon \frac{1}{2c_w} \hat{Z}'_{\mu\nu} \hat{B}^{\mu\nu}
  + \frac{1}{2} M_{Z'}^2 \hat{Z}_\mu^{\prime 2}~,
\eeq
where $c_w$ is the cosine of the Weinberg angle and the hats denote states that
are not mass eigenstates.  After diagonalization one finds a single massless
state, identified to be the photon.  The would-be standard model $Z$ and dark
photon $Z'$ also mix due to electroweak symmetry breaking.  The mixing formulas
can be worked out analytically but are not shown here
(see~\cite{Curtin:2014cca} for the full expressions).

\begin{figure}
\begin{center}
 \includegraphics[width=0.6\textwidth]{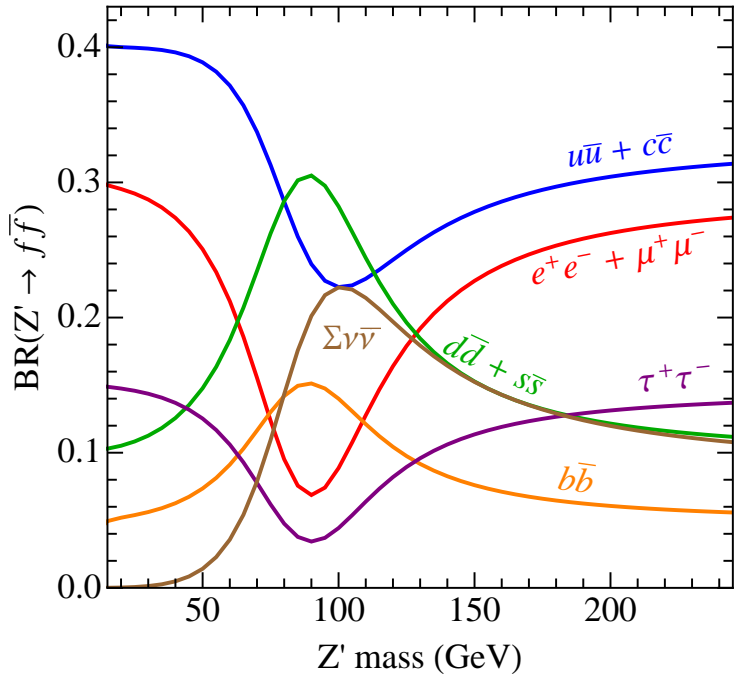}
\end{center}
\caption{Dark photon branching ratios.  These are computed using
$\epsilon = 5 \times 10^{-3}$ although for $\epsilon \ll 1$ the branching
ratios are independent of $\epsilon$.
\label{fig:dphoton_brs}}
\end{figure}

The dark photon inherits couplings to fermions both from mixing with
hypercharge and mixing with the $Z$.  In the limit $\epsilon \ll 1$ and $M_{Z'}
\ll M_Z$ the dark photon couplings to fermions become photon-like and the
partial width simplifies to
\beq
\Gamma(Z' \to f \bar f) = \frac{\alpha M_{Z'}}{3} Q_f^2 N_c \beta_f \left(
 \frac{3 - \beta_f^2}{2} \right) \epsilon^2~,
\eeq
where there are $N_c$ colors of $f$ with charge $Q_f$ and mass $m_f$, and 
\beq
\beta_f^2 \equiv 1 - \frac{4m_f^2}{M^2_{Z'}}~.
\eeq
Ignoring all quark masses except $m_b$ and assuming the top-antitop channel is
closed, the branching ratio of $Z'$ into $\mu^+ \mu^-$ (a convenient and
low-background final state) is
\beq
{\cal B}(Z' \to \mu^+ \mu^-) = 3 \left( 19 + \frac{\beta_b(3-\beta_b^2)}{2}
\right)^{-1}~.
\eeq
When $M_{Z'} \approx M_Z$ the dark photon couplings become $Z$-like and
when $M_{Z'} \gg M_Z$ they become $B$-like.  This can be seen in Fig.\
\ref{fig:dphoton_brs} where we show the branching ratios, assuming the dark
photon decays entirely into standard model particles.  These are computed using
$\epsilon = 5 \times 10^{-3}$ although for $\epsilon \ll 1$ the branching
ratios are independent of $\epsilon$.  For simplicity, we only use the
perturbative calculation.  For low $Z'$ masses, i.e., below a few GeV, it is
necessary to consider threshold effects, QCD corrections, and hadronic
resonances.

\subsubsection{Leptonic production}

One recent use of radiative return has been the search by the KLOE
Collaboration at the DA$\Phi$NE $e^+e^-$ collider for a ``dark photon'' $U$
decaying to $\mu^+ \mu^-$ \cite{Curciarello:2015xja} or $e^+ e^-$
\cite{Palladino:2015jya}, produced by the reaction $e^+ e^- \to \gamma_{\rm
ISR} U$ at an initial CM energy of about 1 GeV.  We make projections for a
similar analysis that can be performed for initial CM energies of 90 and 250
GeV.

To compute the reach we consider the background to be $e^+ e^- \to \gamma
\mu^+ \mu^-$ where the muons come from an intermediate $\gamma^*$ or $Z$.  The
search then proceeds by counting the number of events in the dimuon invariant
mass spectrum.  Since the dark photon width is very narrow the best
significance is achieved by binning as narrowly as possible around the targeted
$Z'$ mass.  The smallest invariant mass bin is determined entirely by the
detector's resolution.  Typically detector resolution for muon-based searches
gets worse at higher momentum (and equivalently higher invariant dimuon mass).
We take the mass resolution to be given $\Delta m = m^2/(10^5~\gev)$ by 
estimating based on the specification $\Delta(1/p_T)=2\times 10^{-5}~\gev^{-1}$
outlined in future detector designs~\cite{Behnke:2013lya}.
\footnote{See Appendix A in~\cite{Reece:2009un} for an estimation of BaBar's
mass resolution.  They find the mass resolution to grow quadratically with
mass.}  For reference, this equates to $\Delta m = 100$ MeV for $m = 100$ GeV.
The limit on $\epsilon$ scales as $(\Delta m)^{-1/4}$ so a 4 times increase in
resolution only results in a 40\% increase in reach on $\epsilon$.

\begin{figure}
\begin{center}
 \includegraphics[width=0.75\textwidth]{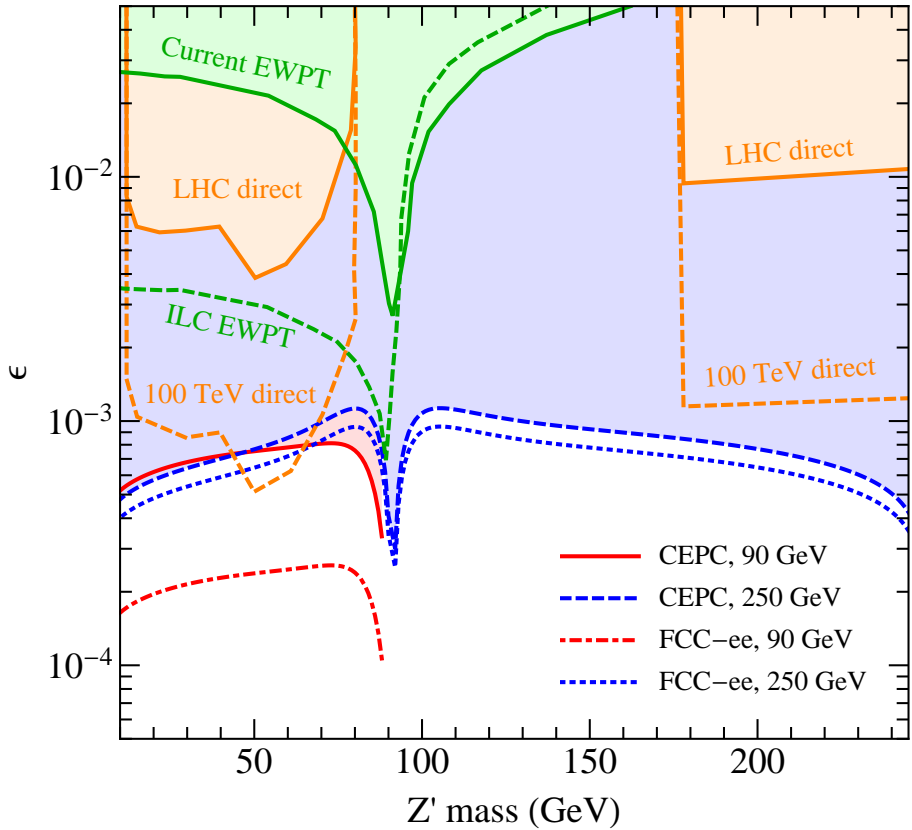}
\end{center}
\caption{Dark photon limits at 95\% C.L. on the hypercharge mixing $\epsilon$
as a function of dark photon mass.  The $\sqrt{s} =$ 90 GeV and 250 GeV lines
show our projections with future $e^+ e^-$ colliders with integrated luminosities
specified in Table~\ref{tab:futurelumis}.  Electroweak precision constraints (EWPT) 
and direct searches are taken from~\cite{Curtin:2014cca}.  The 100 TeV projection 
assumes an integrated luminosity of 3000 fb$^{-1}$.
\label{fig:dphoton_limits}}
\end{figure}

The results are shown in Fig.~\ref{fig:dphoton_limits}.  The current and
projected limits from electroweak precision data (EWPT) were computed
in~\cite{Curtin:2014cca}.  Due to some mild tension in the electroweak fit in
present data~\cite{Baak:2014ora} the inclusion of a dark photon with $M_{Z'} 
> M_Z$ actually improves the fit, which is the reason that projected precision
electroweak limits are weaker.  Alternative projections that assume the
electroweak precision data converges are presented in~\cite{Curtin:2014cca}.
At masses below $M_Z$, the current direct searches are originally taken
from~\cite{Hoenig:2014dsa} which uses the Drell-Yan process $pp \to Z' \to
\ell^+ \ell^-$ normalized to 7 TeV LHC data~\cite{Chatrchyan:2013tia} to
compute limits using the full 7 and 8 TeV LHC data set.\footnote{For current
direct limits on dark photons both above and below $M_Z$ we take limits
from~\cite{Curtin:2014cca} rather than from the original
studies~\cite{Hoenig:2014dsa,Cline:2014dwa}.}  For direct searches for masses
above $M_Z$, the limits are originally taken from~\cite{Cline:2014dwa}, which
recast an ATLAS dilepton search~\cite{ATLAS:2013jma}.\footnote{As far as we can
tell, this limit does not stop at $M_{Z'} \sim$ 175 GeV for any fundamental
reason, but rather
because that is the lowest mass shown in the ATLAS
results~\cite{ATLAS:2013jma}.}.  The 100 TeV direct searches are taken
from~\cite{Curtin:2014cca} which rescaled the previously mentioned direct
limits to 100 TeV with 3000 fb$^{-1}$.

\subsubsection{Hadronic production}

\begin{figure}
\begin{center}
\includegraphics[width=0.6\textwidth]{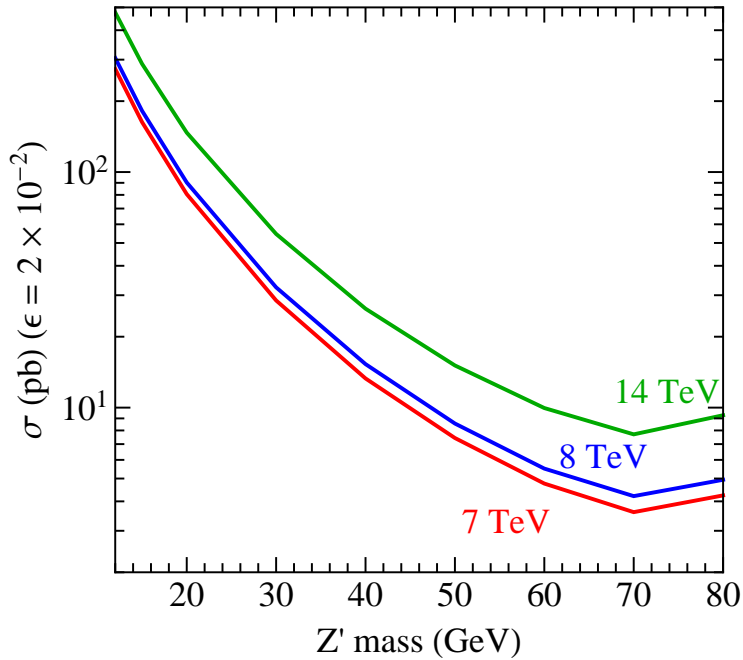}
\end{center}
\caption{Cross sections for $Z'$ production~\cite{Hoenig:2014dsa} based on kinetic mixing at a $pp$ collider with CM energy 7 TeV (red), 8 TeV (blue), and 14 TeV (green).  Plotted values are to be multiplied by $[\epsilon/(2 \times 10^{-2})]^2 = (50 \epsilon)^2$.
\label{fig:haddark}}
\end{figure}

Direct photon production by hadronic collisions in the standard model proceeds
through the subprocess $q \bar q \to \gamma^*$.  Assuming a ``dark photon'' is
produced by  this same process, where the $\gamma$ is now virtual and mixes
kinetically with the dark photon $Z'$, one can utilize Drell-Yan production
of a lepton pair $e^+ e^-$ or $\mu^+ \mu^-$ to evaluate the sensitivity of
dark photon searches in hadronic collisions.  A sample calculation has
been performed in Ref.~\cite{Hoenig:2014dsa} for various LHC energies;
the cross sections are shown in Fig.~\ref{fig:haddark}.  (The reach of a future
100 TeV $pp$ collider has been investigated in Ref.~\cite{Curtin:2014cca}.) 

The peaking of parton distributions at low Feynman $x$ favors low $Z'$ masses.
For example, at 14 TeV a 15 GeV $Z'$ has a cross section of about $300 \times
(2500 \epsilon^2)$ pb $= 750 \epsilon^2$ nb.  Assuming an integrated
luminosity of 1 ab$^{-1}$, this gives rise to at least 10 events when
$\epsilon^2 > 1.3 \times 10^{-11}$.  At such a low $Z'$ mass, however,
background considerations probably dominate any realistic estimate of
sensitivity.

\subsection{Production of $b \bar b$}

\subsubsection{Leptonic production}

The asymmetric $B$ factories PEP-II and KEK-B have explored $b \bar b$
production up to CM energies of about 11 GeV with compelling statistics,
and the upgraded KEK-B with the Belle-II detector will extend samples to
dozens of events per attobarn.  However, from about 11 to 90 GeV the
$e^+ e^-$ territory is much more sparsely populated with data, as one
can see from Table~\ref{tab:colls} and Fig.\ \ref{fig:summary}.  Radiative
return studies from a Giga-$Z$ or Tera-$Z$ factory can help to fill this gap.
A sample process is $e^+ e^- \to (\gamma^*,Z^*) \to b \bar b$, compared for
direct production with the radiative return process $e^+ e^- \to (\gamma^*,Z^*)
\,\gamma_{\rm ISR} \to b \bar b \,\gamma_{\rm ISR}$.

We use lowest-order MadGraph~\cite{Alwall:2011uj} for our estimates of direct and radiative-return
cross sections.  For simplicity we assume that the whole PEP sample of
1.167 events per fb is accumulated at $\sqrt{s} = 29$ GeV, where MadGraph
predicts $\sigma(e^+e^- \to b \bar b) = 36.8$ pb, giving a total sample of
about 43k events.  The corresponding cross sections at 35 and 60 GeV,
relevant for PETRA and TRISTAN, are 25.8 and 16.2 pb, respectively.  With
integrated luminosities of 817 and 942 events per pb (Table~\ref{tab:colls}),
one then has respective samples of 21.1k and 15.3k events from the direct
process at PETRA and TRISTAN.

For radiative return we consider samples integrated over the $\hat E_{CM}$
ranges [10,35], [35,60], and [60,85] GeV, applying Eq.~\eqref{eqn:fl} and
recalling that the beam energy $E$ is $\sqrt{s}/2$.\footnote{In MadGraph the
logarithm is taken to be $\ln(\sqrt{s}/P_{T,{\rm cut}})$; we apply corresponding
corrections of 0.9426 and 0.9471 to the MadGraph results at $\sqrt{s} = 90$
and 250 GeV.}  The results are compared with the direct process in
Table~\ref{tab:bcomp}.

\begin{table}
\caption{Comparison of direct and radiative-return $e^+ e^-$ production of
$b \bar b$.
\label{tab:bcomp}}
\begin{center}
\begin{tabular}{c c c c|c c c c c c} \hline \hline
\multicolumn{4}{c|}{Direct} & \multicolumn{5}{c}{Radiative return} \\
Collider & $\sigma$ & $\int{\cal L}dt$ & Events & $\hat E$ range &
 \multicolumn{2}{c}{$\sqrt{s} =  90$ GeV} &
 \multicolumn{2}{c}{$\sqrt{s} = 250$ GeV} \\
 & (pb) & (pb$^{-1}$) & (10$^3$) & (GeV) & $\sigma$ (pb) & Evts.(10$^6$)$^a$
 & $\sigma$ (pb) & Evts.(10$^3$)$^b$ \\
\hline
PEP      & 36.8 & 1167 & 42.9 & 10--35 & 0.494 & (24.5,0.245)
 & 0.066 & (660,330) \\
PETRA    & 25.8 & 817 & 21.1 & 35--60 & 0.410 & (20.5,0.205)
 & 0.039 & (391,196) \\
TRISTAN  & 16.2 & 942 & 15.3 & 60--85 & 12.94 & (647,6.47)
 & 0.256 & (2562,1281) \\
\hline \hline
\end{tabular}
\end{center}
\leftline{$^a$ Assuming $\int{\cal L}dt = (50, 0.5)$ ab$^{-1}$ at (FCC-ee,
CEPC).}
\leftline{$^b$ Assuming $\int{\cal L}dt = (10, 5)$ ab$^{-1}$ at (FCC-ee,
CEPC).}
\end{table}

Even if all the data from PEP, PETRA, and TRISTAN are pooled, they
are less than the sample that would be gained in the $\hat E_{CM}$
range from 10 to 60 GeV by studying radiative return from an $e^+ e^-$
collider at $\sqrt{s} = 90$ or 250 GeV.

\subsubsection{Hadronic production}

It is not straightforward to compare leptonic and hadronic $b \bar b$
production because the background circumstances are different.  However,
the LHCb Collaboration has demonstrated great sensitivity to specific
final states in which backgrounds can be largely overcome.  As one
example, a recent study of $b$ hadron lifetimes \cite{Aaij:2014owa}
based on a data sample of 1 fb$^{-1}$ at $\sqrt{s} = 7$ TeV accumulates
a sample of $229,439 \pm 503$ $B^+ \to J/\psi K^+$ events, with $J/\psi$
decaying to $\mu^+ \mu^-$.  Given the branching fractions ${\cal B}
(B^+ \to J/\psi K^+) = (1.027 \pm 0.031) \times 10^{-3}$ and ${\cal B}
(J/\psi \to \mu^+ \mu^-) = (5.93 \pm 0.06)\%$ \cite{PDG2014}, this
corresponds to about $(3.77 \times 10^9)/\epsilon_f$ $B^+$ produced,
where $\epsilon_f<1$ is the acceptance for the final state $f$ in question.
In fact, the production cross sections for $B$ mesons at the LHC have
been measured \cite{Aaij:2013noa}:
\bea
\sigma(pp \to B^+ + X) &=& (38.9 \pm 0.3 \pm 2.5 \pm 1.3) \mu{\rm b}~,\\
\sigma(pp \to B^0 + X) &=& (38.1 \pm 0.6 \pm 3.7 \pm 4.7) \mu{\rm b}~,\\
\sigma(pp \to B_s + X) &=& (10.5 \pm 0.2 \pm 0.8 \pm 1.0) \mu{\rm b}~,
\eea
where the errors are statistical, systematic, and normalization (based
on prior branching fraction measurements).  This would correspond to
about $3.9 \times 10^{10}$ $B^+$ in a sample of 1 fb$^{-1}$, yielding
an estimate of $\epsilon_f \simeq 10\%$.

\section{Some accessible questions in heavy flavor spectroscopy \label{sec:flavor}}

Much progress in heavy flavor spectroscopy has been made using the $B$
factories PEP-II and KEK-B.  However, these machines were limited, as will be
the KEK-B upgrade, to CM energies not much above 11 GeV.  There are a number of
questions in the spectroscopy of hadrons containing heavy (charm and bottom)
quarks that could benefit from higher CM energies.  Can an $e^+ e^-$ collider
with CM energy 250 GeV and luminosity $10^{34}$ cm$^{-2}$s$^{-1}$ provide
integrated luminosity to study such states significantly above what has already
been provided by PEP, PETRA, TRISTAN, and LEP?  A sharper answer can be
provided by considering specific processes.

\subsection{Bottomonium analogues of charmonium $X,Y,$ and $Z$ states}

There are a number of charmonium states that appear to contain extra
light quarks or to be admixtures of $c \bar c$ and charmed meson pairs.
Some analogues of these have been seen in the bottomonium sector, but
so far the $X_b$, the analogue of the $X(3872)$, has eluded clearcut 
detection. There is an intriguing possibility that $X_b$
might have already been observed, but identified as $\chi_{b1}(3P)$ which
is close in mass and has the same quantum numbers \cite{Karliner:2014lta}.
Electron-positron collisions with CM energy greater than 11 GeV may be
helpful in resolving this issue and allowing for unambiguous 
identification of $X_b$ and related states.

\subsection{Pair production of narrow $B_{sJ}$ states}

The reaction $e^+ e^- \to B_{sJ} + X$ may be used to look for the
$b$-quark analogue of the very narrow $D_{sJ}$ states seen by BaBar, CLEO 
and Belle \cite{Aubert:2003fg,Besson:2003cp,Abe:2003jk}. The relevant
thresholds are discussed in Subsection D below.

\subsection{Doubly heavy flavor production}
With sufficient CM energy one may study such processes as
\bea
e^+ e^- & \to & b \bar b c \bar c + X~, \nonumber \\
e^+ e^- & \to & b \bar b b \bar b + X~, 
\eea
as a precondition for producing doubly heavy mesons such as $B_c$, $B^*_c$, and
doubly heavy baryons such as $\Xi_{bc} = bcq$, and $\Xi_{bb} = bbq$, where $q$
is a light quark.  Until now the latter have never been clearly observed,
even though it is clear they must exist.  As shown
in Ref.~\cite{Karliner:2014gca}, one must be able to see the (known)
$B_c$ state if one expects to be able to detect $\Xi_{bc}$, so we shall
estimate $B_c$ production by radiative return.  We shall consider the case
$E_{CM} = 90$ GeV, assuming that a circular $e^+ e^-$ collider will spend some
time as a Giga- (or Tera-) $Z$ factory.

The mass of $B_c$ is by now very well known \cite{PDG2014}:  $M(B_c) = 6275.6
\pm 1.1$ MeV.  For optimal $B_c$ production, one probably needs to be above
$B_c^{*+} B_c^{*-}$ threshold, which according to the estimate of Ref.~\cite{Karliner:2014gca}
lies between 12.69 and 12.72 GeV.  The cross section
$\sigma(e^+ e^- \to B_c^{*+} B_c^{*-})$ probably rises sharply near
threshold, in the same manner as $\sigma(e^+ e^- \to D^{*+} D^{*-})$, and
may be estimated as follows.

The cross section for $e^+ e^-$ production of a $b \bar b$ pair, far enough
above the $\Upsilon(4S)$, is expected to be
\beq \label{eqn:sub}
\sigma(e^+ e^- \to b \bar b; \hat s) = \frac{4 \pi \alpha^2}{3 \hat s}
\cdot \frac{1}{3} = \frac{4 \pi \alpha^2}{3 s} \cdot \frac{1}{x} \cdot
\frac{1}{3}~,
\eeq
and is about 180 pb at $\sqrt{\hat s} = 12.72$ GeV.  By comparing cross
sections for $B^+$ and $B_c$ production at LHCb, Ref.~\cite{Karliner:2014gca}
found the probability of a $b$ quark fragmenting to $B_c^- = b \bar c$ to be
about $10^{-2}$.  Thus near $B_c^{*+} B_c^{*-}$ threshold, one might expect
\beq
\sigma(e^+ e^- \to B_c^+ B_c^- + X) \simeq 1.8 ~{\rm pb}~,
\eeq
where $X$ denotes the possibility of one or two additional photons from
$B_c^*$ decays.

The cross section, Eq.~\eqref{eqn:sub}, may now be multiplied by $2 f_e(x,\sqrt{s},
p_{T,{\rm cut}})$ (see Eq.~\eqref{eqn:split}) and integrated over an
appropriate range of $x$.  The $B_c^*$ form factor and the $1/\hat s$ factor in
the cross section will introduce some suppression, which we shall imitate by
introducing a maximum $\hat s_{\rm max} = (20 ~{\rm GeV})^2$.  For $E_{CM} =
90$ GeV, we thus perform the integral
\beq \label{eqn:bc}
\sigma(e^+ e^- \to \gamma B_c^+ B_c^- + X) = \frac{2 \alpha}{\pi}~\ln \frac{E}
{m_e}~\int \frac{dx}{x}~(35.7~{\rm fb}) = 1.7~{\rm fb}~.
\eeq
Here we have neglected the small deviation of $(1+x^2)/(1-x)$ from 1, taken the
limits of integration between $x_{\rm min} = (12.72/90)^2$ and $x_{\rm max} =
(20/90)^2$, and used $E = 45$ GeV.  One still has to pay the price of the $B_c$
branching fraction to an observable final state, but as we expect
${\cal B}(B_c \to J/\psi \mu \nu)$ to exceed a percent \cite{Karliner:2014gca}
this seems possible with a sample exceeding one event per ab.

At a different center-of-mass energy, as long as the same range of $\sqrt{\hat
s}$ is taken, one can show that the cross section in Eq.~\eqref{eqn:bc} scales
as $(1/s)\ln(E/m_e)$, so at $\sqrt{s} = 250$ GeV, it becomes 0.24 fb.

\subsection{Interesting thresholds}

The production of $B \bar B$ pairs has occupied most of the running time of
the $B$ factories KEK-B and PEP-II.  However, some data have been taken at
higher energies, as indicated in Table~\ref{tab:colls}.  The CLEO Collaboration
has taken a small amount of data above $\Lambda_b \bar \Lambda_b$ threshold
in search of a ``magic energy'' for $\Lambda_b$ pair production; none was
found.  In Table~\ref{tab:thr} we summarize some thresholds for heavy flavor
production in $e^+ e^-$ collisions.

\begin{table}
\caption{Some thresholds for heavy flavor production in $e^+ e^-$ collisions.
\label{tab:thr}}
\begin{center}
\begin{tabular}{c c} \hline \hline
      Final state   & Threshold \\
                    &   (MeV)   \\ \hline
      $B \bar B$    & 10559 \\
     $B \bar B^*$   & 10605 \\
    $B^* \bar B^*$  & 10650 \\
    $B_s \bar B_s$  & 10734 \\
   $B_s \bar B^*_s$ & 10782 \\
 $B^*_s \bar B^*_s$ & 10831 \\
 $B_{s0} \bar B_s^*$  & 11132--$11193^a$ \\
$\Lambda_b \bar \Lambda_b$ & 11239 \\
   $B_c \bar B_c$   & 12551 \\
  $B_c \bar B^*_c$  & 12619--12635$^b$ \\
 $B^*_c \bar B^*_c$ & 12687--12719$^b$ \\
 $\Xi_{bc} \,\bar \Xi_{bc}$ & 13842--13890$^c$ \\
 $\Xi_{bb} \,\bar \Xi_{bb}$ & 20300--20348$^c$ \\ \hline \hline
\end{tabular}
\end{center}
\leftline{$^a$See text. $^b$With estimated $B_c^*$--$B_c$ splitting 68--84
MeV \cite{Karliner:2014gca}. $^c$ Estimate in \cite{Karliner:2014gca}.}
\end{table}

Here we have used masses tabulated in Ref.~\cite{PDG2014}.  The state $B_{s0}$
in Table~\ref{tab:thr} is the expected analogue, with $J^P = 0^+$, of the
$D_{s0}(2317)$, which is narrow because it lies below $D K$ threshold.
In order to produce the $B_{s0}$ in $e^+ e^-$ collisions, it must be
accompanied by a $\bar B^*_s$ or heavier companion. Angular momentum and parity
conservation forbid the process $e^+ e^- \to \gamma^* \to B_{s0} \bar B_s$.
The $B_{s0}$ mass is estimated to be 5717 MeV by assuming that $D_{s0}$
and $D_{s0}$ are chiral partners of $D_s$ and $B_s$ and therefore the
$B_{s0}-B_s$ splitting is very close to the $D_{s0}-D_s$ splitting
\cite{Bardeen:2003kt,Nowak:2003ra}. On the other hand, in order for
$B_{s0}$ to be interesting, it needs to be narrow. In analogy with $D_{s0}$ 
which is narrow because it is below the $D K$ threshold,
$B_{s0}$ needs to be below $B K$ threshold, i.e.,
below 5778 MeV. So in any case the interesting threshold is between 
5717 MeV + $m_{B_s^*}$ and 5778 MeV + $m_{B_s^*}$, i.e., between
11132 MeV and 11193 MeV.

\section{Conclusions \label{sec:conclusions}}

While $e^+ e^-$ collisions have been studied with impressive statistical power
at energies accessible to the asymmetric $B$ factories KEK-B and PEP-II, the
CM energy range from about 12 to 80 GeV accessible to PEP, PETRA, and TRISTAN
is much less thoroughly investigated. The radiative return process $e^+ e^- \to
\gamma_{\rm ISR} e^+ e^- \to \gamma_{\rm ISR} f$, where ISR denotes
initial-state radiation, can help fill this gap.  Some examples are given of
processes that could be investigated using radiative return, starting from
a collider operating at 90 or 250 GeV.  Although the same final states $f$ can
often be produced with higher cross sections in hadronic collisions, the
relative cleanliness of the $e^+ e^-$ environment gives it an advantage whose
quantitative value must be investigated using detailed detector simulation.

Processes which could benefit from radiative return studies at high energies
include searches for ``dark photons'' $Z'$, heavy quark (particularly $b$)
production, and spectroscopy of states too heavy to be produced at the
asymmetric $B$ factories.  In studying subenergies in the 12--80 GeV range, it
was found advantageous to use total $e^+ e^-$ CM energies near the $Z$ rather
than at the highest possible energy.

\section*{Acknowledgements}

We thank Gideon Alexander, Henryk Czyz, Achim Denig, Stefania Gori, David
Tucker-Smith, Graziano Venanzoni, and Sau Lan Wu for helpful discussions.  The
work of J.L.R. was supported in part by the U.S. Department of Energy, Division
of High Energy Physics, Grant No.\ DE-FG02-13ER41958, and by funds from the
Physics Department of the University of Chicago.  L.-T.W. is supported by DOE
grant DE-SC0003930.  Monte Carlo computations were performed on the Midway
cluster supported by the Research Computing Center at the University of Chicago.

\section{Appendix: Parton luminosity \label{sec:plumi}}

Another way to understand the rate for a process below the nominal
center-of-mass energy is using parton luminosities \cite{Martin:2009iq,%
Quigg:2009gg}.  In this parameterization the cross section is written as
\bea
\sigma(s) & = & \int d\tau \frac{dL_{ab}}{d\tau} \hat{\sigma}_{ab}(\hat{s})~, \nonumber \\
          & = & \int \frac{d\tau}{\tau} \left(\frac{1}{s}\frac{dL_{ab}}
          {d\tau}\right) [\hat{s} \hat{\sigma}_{ab}(\hat{s})]~,
  \label{eqn:plum}
\eea
where $\tau = \hat{s}/s$ and $a,b$ specify the incoming parton species.  The hatted quantities are with respect to the colliding partons, primarily electron and photons for $e^+ e^-$ colliders and quarks and gluons for $pp$ or $p\bar{p}$ colliders.  The quantity $(1/s)(dL_{ab}/d\tau)$ is called the {\it parton luminosity} and has units of a cross section.  It is computed as
\bea
  \frac{d L_{ab}}{d\tau}
  = \frac{1}{1+\delta_{ab}} \int_\tau^1 \frac{dx}{x} \left[ f_a(x) f_b
 \left(\frac{\tau}{x}\right) + f_a\left(\frac{\tau}{x}\right) f_b(x) \right]~.
\eea
In the following subsections we show the parton luminosity calculated for lepton colliders and hadron colliders as a means to compare the expected rates at the different types of colliders.  An important caveat is to note that there are significant differences to actually computing rates at the different machines which means just comparing the parton luminosity values can give an inaccurate picture.  More realistically one needs to consider leptonic versus hadronic branching ratios, detector efficiencies for the final states, and background processes.

\subsection{Leptonic parton luminosity}

At an $e^+ e^-$ collider when considering the only initial state as $e^+ e^-$
and not $e^+ \gamma$, $e^- \gamma$, or $\gamma\gamma$, the electron
distribution function is almost the same as the splitting function $f_e(x)$
in Eq.~\eqref{eqn:split}:
\beq
  f_e(x,s) = \delta(1-x) + \frac{\alpha}{\pi} \ln\frac{E}{m_e}
  \left( \frac{1+x^2}{(1-x)_+} + \frac{3}{2}\delta(1-x) \right)
  + \mathcal{O}(\alpha^2)~.
\eeq
The plus distribution regularizes the behavior at $x=1$ \cite{Altarelli:1977zs} and the $(3/2)\delta(1-x)$ is for overall normalization.  The inclusion of the $\delta$-function is necessary because the distribution function is inclusive and needs to account for the no-splitting case.

\begin{figure}
\begin{center}
  \includegraphics[width=0.48\textwidth]{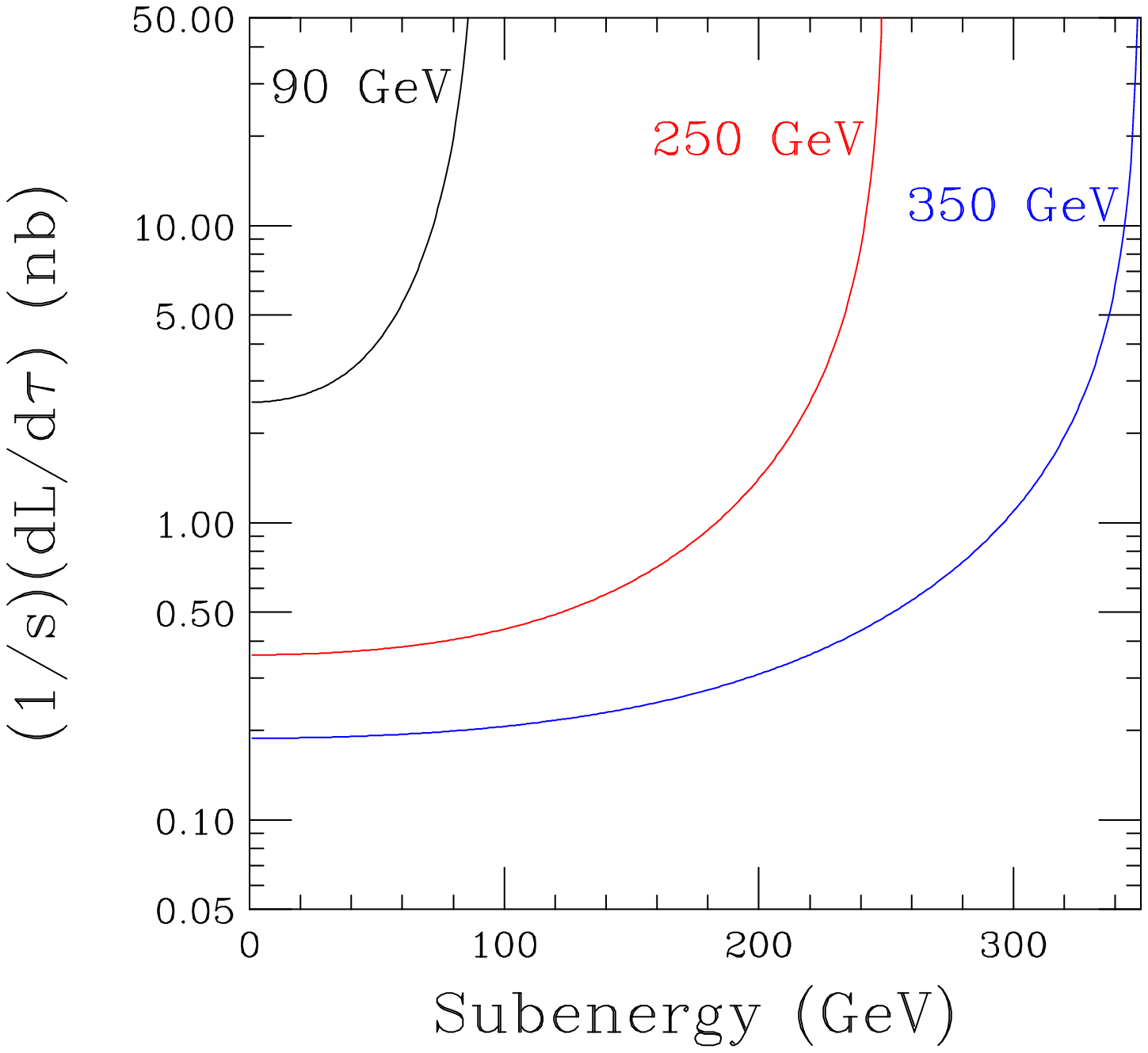}
  \includegraphics[width=0.48\textwidth]{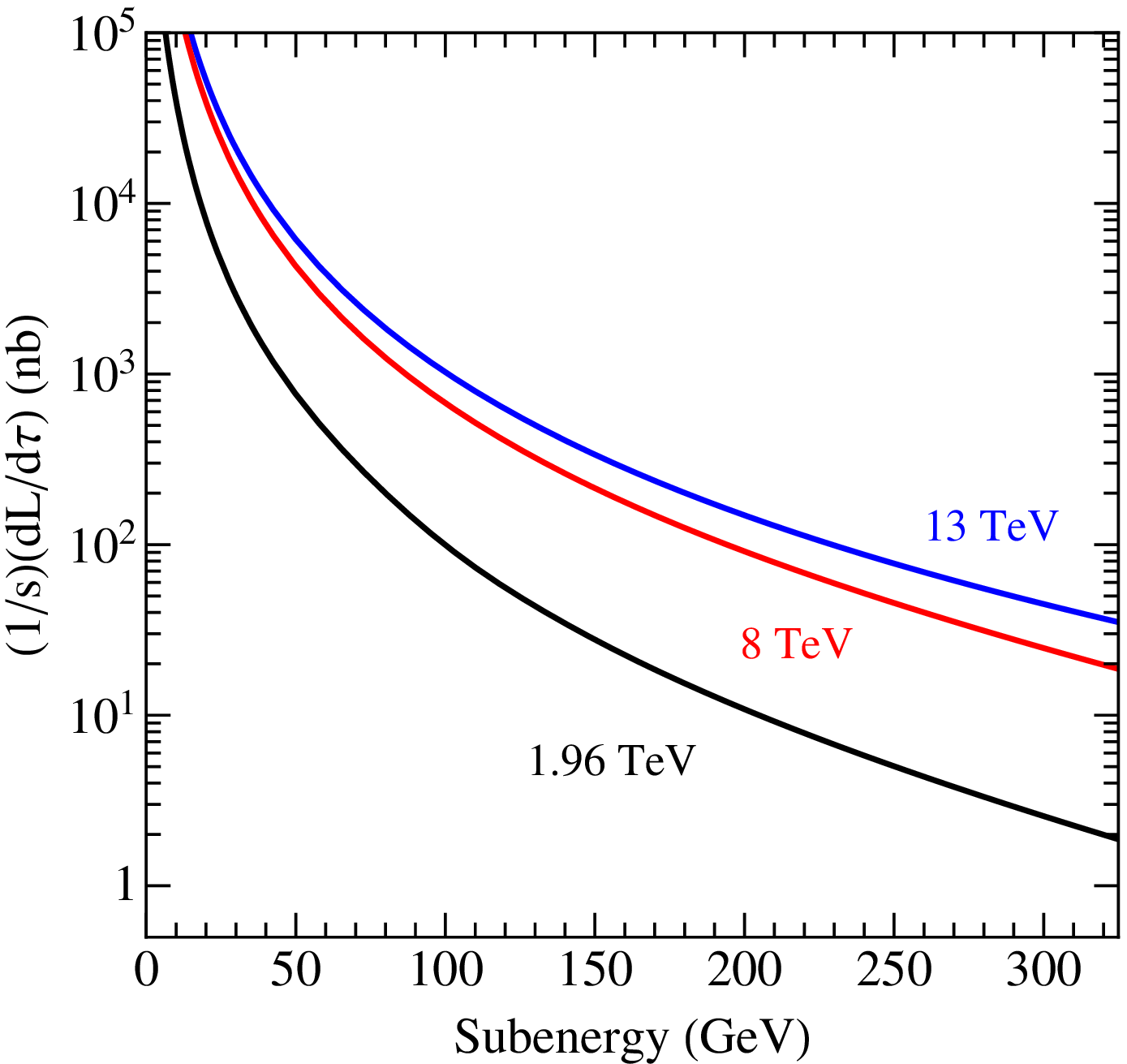}
  \caption{Parton luminosities for $e^+ e^-$ at $\sqrt{s} = 90~\gev, 250~\gev, 350~\gev$ (left) and for $p\bar{p}$ at $\sqrt{s} = 1.96~\tev$ and $pp$ at $\sqrt{s} = 8~\tev, 13~\tev$ (right).}
  \label{fig:plumi}
\end{center}
\end{figure}

For the parton luminosity we find
\beq
  \frac{dL}{d\tau}=
  \delta(1-\tau)\left(1+\frac{3\alpha}{\pi} \ln \frac{E}{m_e} +
  \mathcal{O}(\alpha^2) \right) + \frac{2\alpha}{\pi} \ln \frac{E}{m_e}
  \frac{1+\tau^2}{1-\tau} + \mathcal{O}(\alpha^2) .
\eeq
Fig.~\ref{fig:plumi} (left) shows the parton luminosity for several different CM energies.  For narrow resonances, we can directly use the parton luminosity to compute rates.  According to the narrow width approximation for a resonance of spin $J$, mass $m$, and total width $\Gamma$
\beq
  \hat{s} \hat{\sigma}(\hat{s}) =
  4 \pi^2 (2J+1) \bri \brf m \Gamma \delta(\hat{s} - m^2)~,
\eeq
giving a cross section of
\beq
  \sigma(s) =
  4 \pi^2 (2J+1) \brf \frac{\Gamma_i}{m}
  \left.\left(\frac{1}{s}\frac{dL}{d\tau}\right)\right|_{\tau=m^2/s}~.
\eeq
We can evaluate the appropriate $s$ curve in Fig.~\ref{fig:plumi} at a
subenergy $m$ and find the rate by multiplying by $\Gamma/m$, a spin factor,
and branching ratios.  As an example, with $\Gamma_{ee} = 0.322$ keV and $m =
10.5794$ GeV, we find $\sigma(e^+e^- \to \gamma \Upsilon(4S);
s=(90~{\rm GeV})^2) = 9.17$ fb, the value obtained in Section~\ref{sec:resonance}.

\subsection{Hadronic parton luminosity}

For comparison we consider a hadron collider with the initial state
\beq
q\bar{q} = \{ u\bar{u}, d\bar{d}, s\bar{s}, c\bar{c}, \bar{u}u, \bar{d}d, \bar{s}s, \bar{c}c \}.
\eeq
The parton distribution functions for the proton are non-perturbative functions
that describe the probability a given parton is taken from the proton.

Fig.~\ref{fig:plumi} (right) shows the parton luminosity for several different
CM energies.  In contrast to lepton colliders, a higher-energy hadron collider
also increases the rates at lower subenergies.

The parton distributions are accessed via the LHAPDF interface
\cite{Butterworth:2014efa}. The sets used are \texttt{CT10nnlo\_as\_0118}
set~\cite{Gao:2013xoa}, \texttt{MTSW2008nnlo68cl}~\cite{Martin:2009iq}, and
\texttt{NNPDF23\_nnlo\_as\_0118}~\cite{Ball:2012cx}.  Only the \texttt{CT10}
set is shown in Fig.~\ref{fig:plumi} as the differences are negligible in the
figure.

\section{Appendix: Fractional luminosity for $e^+ e^- \to \mu^+ \mu^-$
\label{sec:flumi}}

In Section~\ref{sec:continuum} we discussed fractional luminosity using factorization in the collinear limit to compute $\sigma(s) \equiv \sigma(e^+ e^- \to \gamma f;s)$ in terms of $\sigma(e^+ e^- \to f;s)$ with the results shown in Fig.~\ref{fig:frac}.  The use of factorization is very convenient to obtain general results independent of the final state $f$.

\begin{figure}
\begin{center}
  \includegraphics[width=0.48\textwidth]{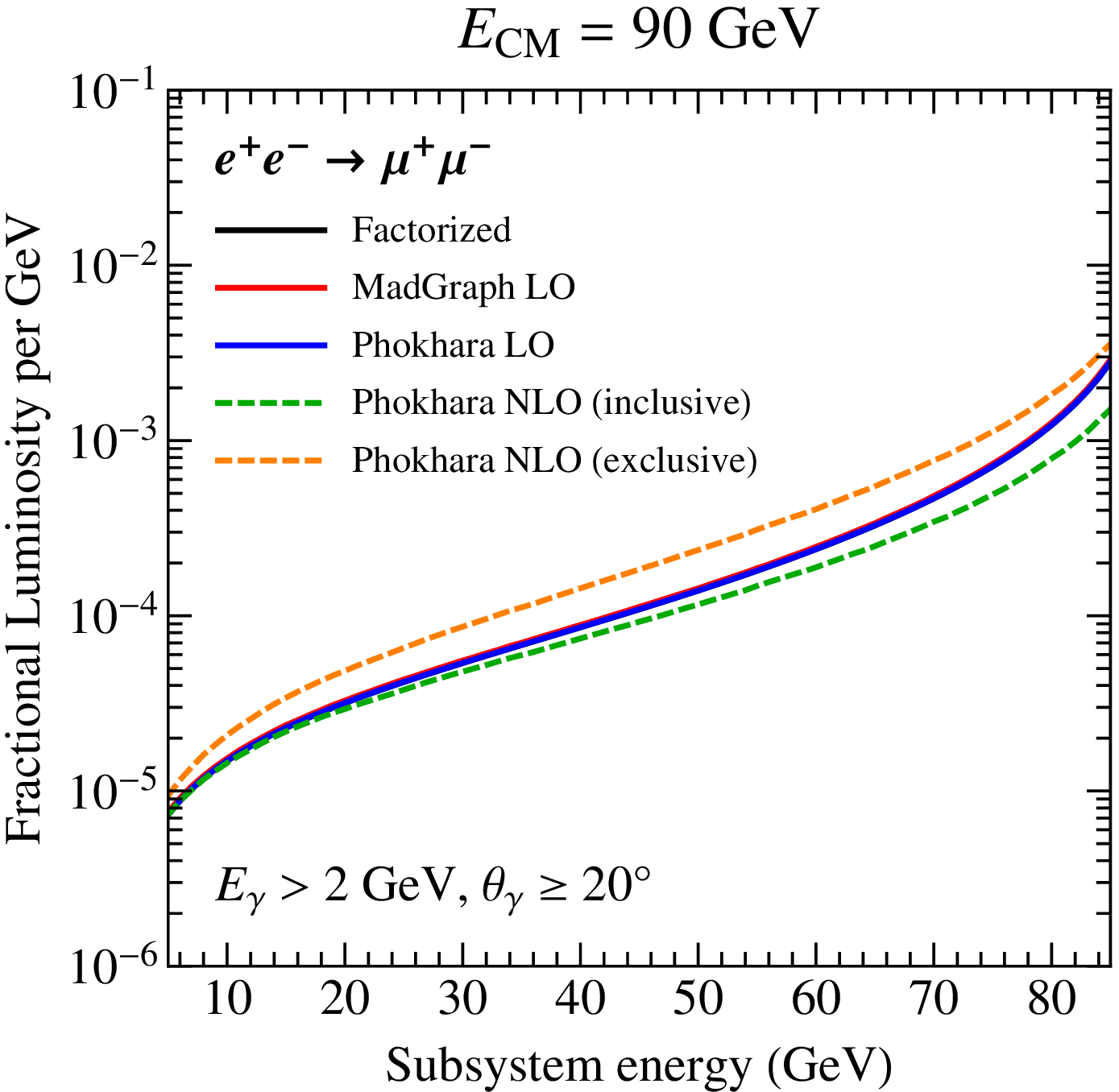}
  \includegraphics[width=0.48\textwidth]{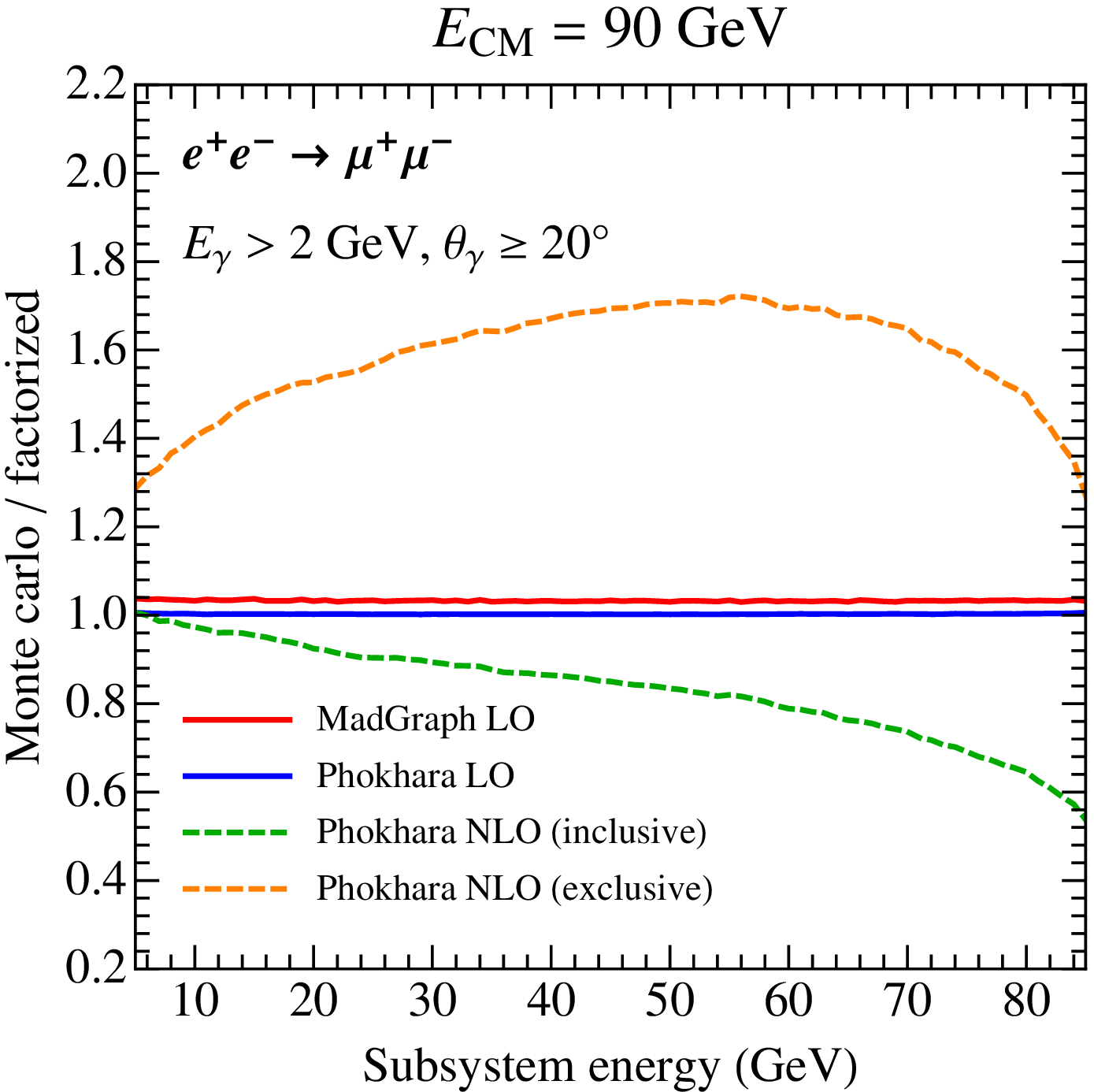}
  \caption{Fractional luminosity for process $e^+ e^- \to \mu^+ \mu^-$ at
$\sqrt{s} = 90$ GeV with an angular cut of $20^\circ$ (left) and showing the
difference between the Monte Carlo and analytic calculation (right).  A bin size of $\Delta = 1$ GeV is used.}
  \label{fig:flumi90mumu}
\end{center}
\end{figure}

Here we compute the fractional luminosity exactly with the processes $\sigma(e^+ e^- \to \mu^+ \mu^-;s)$ and $\sigma(e^+ e^- \to \gamma \mu^+ \mu^-;s)$ to demonstrate that the factorized form used in the main text was justified.  We use the Monte Carlo programs MadGraph~\cite{Alwall:2011uj} and Phokhara~\cite{Campanario:2013uea}.  Fig.~\ref{fig:flumi90mumu} (left) shows the fractional luminosity compared to our computation from Section~\ref{sec:continuum} for $\theta_\gamma \geq 20^\circ$ and $E_\gamma > 2$ GeV.  The energy cutoff is necessary to cut off the soft divergence.  Fig.~\ref{fig:flumi90mumu} (right) shows the ratio of the Monte Carlo to the analytic factorized form.  The Phokhara result agrees well with the factorized form and the MadGraph results differ in normalization by $\approx 5\%$.  Results for other angular cuts are similar.

Fig.~\ref{fig:flumi90mumu} also displays next-to-leading (NLO) order results computed with Phokhara.  Interpreting these results requires some additional explanation.  Recall the definition of fractional luminosity
\beq \label{eqn:flumi}
\frac{d \sigma(s)}{d \hat E_{CM}} \Delta \equiv \LLf \hat \sigma(\hat s)~,
\eeq
where $\sigma(s)$ is the three-body cross section and $\hat{\sigma}(\hat{s})$ is the two-body cross section.

Explicitly, the left side of Eq.~\eqref{eqn:flumi} evaluates the differential three-body distribution at a given $\hat{E}_{CM}$ value and is multiplied by the bin width $\Delta$ to get a three-body cross section (i.e., one term of a Riemann sum).  The right side of Eq.~\eqref{eqn:flumi} is the two-body cross section evaluated at $\hat{E}_{CM}^2 = \hat{s}$ with a coefficient identified as the fractional luminosity.

At leading order (LO), this definition is unambiguous because the two-body
cross section is a $\delta$-function in $\hat{s}$.  At NLO, however, the
two-body cross section becomes a distribution in $\hat{s}$ due to real photon
emission.\footnote{Considering real emission an $N$-body process becomes an
$N+1$-body process, but for simplicity we will label it by its LO phase space,
so that even with real emission it is called an $N$-body process.}  One needs
to choose whether to define the two-body cross section as the integral over all
$\hat{s}$ values or the integral over a small window $\Delta$ near the nominal
value.

The former we call the {\it inclusive} NLO cross section and evaluate as
\beq
\hat{\sigma}(\hat{E}_{CM}) = 
\int^{\hat{E}_{CM}}_{0\quad\quad} \left(\frac{d\hat{\sigma}}{d\hat{E}'_{CM}}\right) d\hat{E}'_{CM}~,
\eeq
and the latter we call the {\it exclusive} NLO cross section
\beq
\hat{\sigma}(\hat{E}_{CM}) = 
\int^{\hat{E}_{CM}}_{\hat{E}_{CM} - \Delta}  \left(\frac{d\hat{\sigma}}{d\hat{E}'_{CM}}\right) d\hat{E}'_{CM}~.
\eeq
Both versions are shown in Fig.~\ref{fig:flumi90mumu} in green (lower) and orange (upper) dashed lines, respectively.

The inclusive result has a lower fractional luminosity because the two-body
cross section is larger than the LO result.  This is because the real photon
emission already induces some radiative return.  The decrease of the fractional
luminosity, relative to LO, at high subenergies is due to the three-body cross
section decreasing because of the 2 GeV cut on photon energy.

On the other hand, the exclusive result has a lower two-body cross section because the integration only includes values of $\hat{E}_{CM}$ near the nominal value while the phase space at lower values is still populated.  The three-body cross section is the same as the inclusive case so the same decrease at high subenergies is observed.  The exclusive result depends on the bin size used.  In Fig.~\ref{fig:flumi90mumu} we show results using a bin size of $\Delta = 1$ GeV.  As the bin size is increased, the exclusive result approaches the inclusive result.

\end{document}